\documentclass[manuscript,screen]{acmart}

\setcopyright{none}         

\usepackage{amssymb}
\usepackage{amsmath,amsfonts}
\usepackage{algorithmic}
\usepackage{graphicx}
\usepackage{subcaption}     
\usepackage{enumitem}
\usepackage{tabularx}
\usepackage{threeparttable}
\usepackage{xcolor}
\definecolor{grey}{gray}{0.9}
\usepackage{xspace}
\usepackage{hyphenat}
\usepackage{booktabs}       
\usepackage{soul}
\usepackage{hyperref}
\usepackage{wrapfig}
\usepackage{multirow}
\usepackage{listings}
\usepackage{fancybox}
\usepackage{enumitem}
\usepackage{graphicx}
\usepackage{color}
\usepackage{xcolor}
\usepackage{fancyvrb}
\usepackage{array}
\usepackage{amsmath}
\usepackage{xspace}
\usepackage{url}
\usepackage{tikz}
\usepackage{caption}
\usepackage{pgfplots}
\usepackage{balance}
\usepackage{subcaption}
\pgfplotsset{compat=1.16}
\usepackage{pgf-pie}
\usepackage{float}
\usetikzlibrary{pgfplots.statistics,calc}
\usepackage{algorithm}
\usepackage{algorithmic}
\usepackage{adjustbox}
\usepackage{makecell}

\usepackage{tcolorbox}

\definecolor{songcolor}{RGB}{191,191,191}

\usepackage{xcolor}
\newcommand{\song}[1]{}
\newcommand{\jiho}[1]{}

\newcommand{\respto}[1]{}

\newcommand{\citeresp}[1]{}

\usepackage{comment}
\usepackage{tcolorbox}
\usepackage{etoolbox}

\newcommand{\yinghang}[1]{}
\newcommand{\Foutse}[1]{}
\newcommand{\jack}[1]{}
\newcommand{\shinhwei}[1]{}
\newcommand{\leuson}[1]{}
\newcommand{\defectj}{\textsc{Defects4J}\xspace}
\newcommand{\SWE}{\textsc{SWE-bench Lite}\xspace}
\newcommand{\SWEA}{\textsc{SWE-agent}\xspace}
\newcommand{\tool}{\textsc{BloomAPR}\xspace}
\newcommand{\nolinefrac}[2]{\genfrac{}{}{0pt}{}{#1}{#2}}
\newcommand{\todo}[1]{\textcolor{red}{{\it [TODO: #1]}}}
\definecolor{darkred}{RGB}{139,0,0}
\newcommand{\add}[1]{{#1}}

\makeatletter
\newcommand{\mybox}[1]{%
    \setbox0=\hbox{#1}%
    \setlength{\@tempdima}{\dimexpr\wd0+13pt}%
    \begin{tcolorbox}[boxrule=1.0pt, colback=grey, arc=4pt,
        left=2pt,right=2pt,top=2pt,bottom=2pt,boxsep=2pt]
        #1
    \end{tcolorbox}
}
\makeatother

\begin{document}

\newpage
\title{BloomAPR: A Bloom's Taxonomy-based Framework for Assessing the Capabilities of LLM-Powered APR Solutions}

\author{Yinghang Ma}
\affiliation{%
  \institution{York University}
  \city{Toronto}
  \country{Canada}}
\email{coppelia@cse.yorku.ca}

\author{Jiho Shin}
\affiliation{%
  \institution{York University}
  \city{Toronto}
  \country{Canada}}
\email{jihoshin@yorku.ca}

\author{Leuson Da Silva}
\affiliation{%
  \institution{Polytechnique Montréal}
  \city{Montreal}
  \country{Canada}}
\email{leuson-mario-pedro.da-silva@polymtl.ca}

\author{Zhen Ming (Jack) Jiang}
\affiliation{%
  \institution{York University}
  \city{Toronto}
  \country{Canada}}
\email{zmjiang@cse.yorku.ca}

\author{Song Wang}
\affiliation{%
  \institution{York University}
  \city{Toronto}
  \country{Canada}}
\email{wangsong@yorku.ca}

\author{Foutse Khomh}
\affiliation{%
  \institution{Polytechnique Montréal}
  \city{Montreal}
  \country{Canada}}
\email{foutse.khomh@polymtl.ca}

\author{Shin Hwei Tan}
\affiliation{%
  \institution{Concordia University}
  \city{Montreal}
  \country{Canada}}
\email{shinhwei.tan@concordia.ca}


\begin{abstract}

\add{
Recent advances in large language models (LLMs) have accelerated the development of AI-driven automated program repair (APR) systems. However, these systems are still typically evaluated on static benchmarks such as \defectj and SWE-bench, which suffer from two key limitations: (1) the risk of data contamination, potentially inflating reported performance through overlap with LLM training data, and (2) limited ability to assess APR capabilities under increasingly diverse repair conditions. In this paper, we introduce \tool, a dynamic evaluation framework grounded in Bloom's Taxonomy. Our framework provides a structured way to assess the capabilities of LLM-powered APR systems across progressively more demanding reasoning layers. Using \defectj as a case study, we evaluate two representative LLM-powered APR systems, ChatRepair and CigaR, across six LLMs: GPT-3.5-Turbo, Llama-3.1, StarCoder-2, GPT-5.4 mini, Qwen3.5-Plus, and Qwen3-Coder. We report three evaluation metrics throughout the study: \emph{Plausible Patch (PP)}, \emph{SYntactic Equivalence (SYE)}, and \emph{Exact Match (EM)}. Our results show that these systems achieve a PP rate of up to 82.95\% on the memory-permissive \texttt{Remember} layer. Under the \texttt{Understand} layer, identifier-level perturbations substantially reduce PP, which drops to 21.66\%–57.60\% under \texttt{Hash-based renaming} and 28.11\%–55.76\% under \texttt{Rephrasing-based renaming}. At the \texttt{Apply} layer, PP reaches 72.35\%–94.47\% on lightweight, single-file synthetic cases, whereas SYE and EM remain considerably lower, at 18.43\%–34.10\% and 17.51\%–33.64\%, respectively. Finally, in the \texttt{Analyze} layer, where the same underlying root causes must be repaired in substantially different real-world project environments, performance remains limited: on the finalized 104-instance Analyze dataset, which contains one validated real-world instance for each of 104 matched \defectj bug IDs across 28 real-world projects, PP ranges from 2.88\% to 23.08\%, while SYE and EM both range from 0.00\% to 10.58\%. A second case study on \SWE using 190 selected single-function cases shows a similar layered pattern. These two case studies provide preliminary cross-benchmark evidence that the four-layer trend is not specific to a single static APR benchmark, even though exact quantitative values vary with the benchmark and evaluated agent. These findings highlight a clear gap between benchmark-level success and robust, context-sensitive repair, underscoring the need for dynamically evolving benchmarks and more trustworthy evaluation of LLM-powered software engineering systems.\respto{R1-1.5}}

\end{abstract}




\maketitle

\section{Introduction}
\label{intro}

\add{Artificial Intelligence (AI) has recently gained rapid popularity in both research and practice. This surge is primarily driven by the increasing adoption of Large Language Models (LLMs), which have unlocked new possibilities and are poised to transform our society significantly~\cite{GPTJobMarket2024, Partington2024}. 
Take software engineering (SE), for example. LLMs have shown great promise and are expected to revolutionize the entire software engineering lifecycle~\cite{HouTOSEM2024, 10.1109/TSE.2024.3368208, liu2024largelanguagemodelbasedagents}.
Companies and SE professionals have already begun adopting LLM-powered tools to boost their productivity~\cite{GoogleAICode,10.1145/3663529.3663836} and enhance software quality~\cite{YangASPLOS2025, frommgen2024resolving, wadhwa2024core}. The research progress (e.g.,~\cite{openai2024gpt4technicalreport,10.1109/ICSE48619.2023.00194,10.1145/3650212.3680323,yang2024sweagent}) in this area is primarily evaluated using publicly available benchmarks (e.g., 
Defects4J~\cite{Defects4J}, HumanEval~\cite{rm1}, and SWE-Bench~\cite{jimenez2024swebench}).\respto{E0-3.3} \respto{R1-13.1}}

Although the use of these benchmarks facilitates cross-comparison and replication, the reported results can suffer from data contamination due to the overlap between the training and the evaluation datasets~\cite{ramos2025largelanguagemodelsmemorizing, zhang2025UnseenHorizons}. Consequently, the benchmark results may neither accurately reflect true model capabilities nor reveal their deficiencies~\cite{xia2024leaderboardrankingcoding, mirzadeh2025gsmsymbolic}. However, faithfully evaluating the capabilities of LLM-powered SE systems presents the following challenges: 

\begin{itemize}[leftmargin=*]
    \item \textbf{The LLM side}: There are over $270$ existing SE benchmarks~\cite{cao2025buildbenchmarkrevisiting274}, with new ones continuously curated or updated based on real-world datasets~\cite{RepoCod, ComplexCodeEval}. Although data decontamination procedures can be applied during LLM training to deliberately search for and remove benchmark datasets from the training corpus~\cite{starcoder2stackv2, llama3herdmodels}, it remains challenging to comprehensively identify and exclude all benchmark data and their associated tasks from the pre-training datasets to ensure trustworthiness. 

    \item \textbf{The benchmark side}: Existing benchmarks are used not only for evaluation and cross-comparison~\cite{OuyangISSTA2024} but also for performance tuning and debugging during research and tool development~\cite{10.1145/3650212.3680323, hidvégi2024cigarcostefficientprogramrepair, alpharepair}. 
    This dual use makes it difficult to keep benchmarks open and reusable for the community while simultaneously avoiding contamination, which is crucial for faithful evaluation.
    Recently, dynamic benchmarking techniques have been proposed to continuously curate fresh data~\cite{livecodebench, RepoCod, ComplexCodeEval, aleithan2024swebenchenhancedcodingbenchmark,majdinasab2025} to minimize the overlap between the training and the evaluation datasets.
    However, newly curated datasets may not preserve the same characteristics (e.g., difficulty levels and skill sets) as the original benchmarks, making it difficult to accurately track the progression of a model's capabilities~\cite{JiangICSE2023}.
        
    \item \textbf{The LLM-powered systems side}: Studies show that LLM-powered SE systems, such as Automated Program Repair (APR) systems, can be brittle, even in response to small code perturbations~\cite{WuISSTA2023, zhang2025UnseenHorizons, wang2022recoderobustnessevaluationcode}. Unfortunately, many existing evaluations assess LLM-powered systems only on benchmark instances as they are, overlooking other aspects such as robustness and generalization. One exception is~\cite{OuyangISSTA2024}, which employs basic mutation techniques (e.g., operator and literal mutations) to assess the robustness of LLM-powered APR systems. Since such synthetic mutations may not accurately capture the complexity of real-world bugs~\cite{10.1145/2635868.2635929, DAKHEL2024107468}, effective bug-fixing capability requires progressively deeper cognitive abilities: recognizing bug contexts and familiar fixing patterns, comprehending underlying buggy behaviors, applying repair knowledge to new instances, and ultimately analyzing root causes across diverse bug manifestations and project contexts. Therefore, novel evaluation methodologies are needed to more comprehensively and realistically assess the capabilities of LLM-powered APR systems.

\end{itemize}

\begin{figure*}[!htbp]
    \centering
    \includegraphics[width=1.0\textwidth]{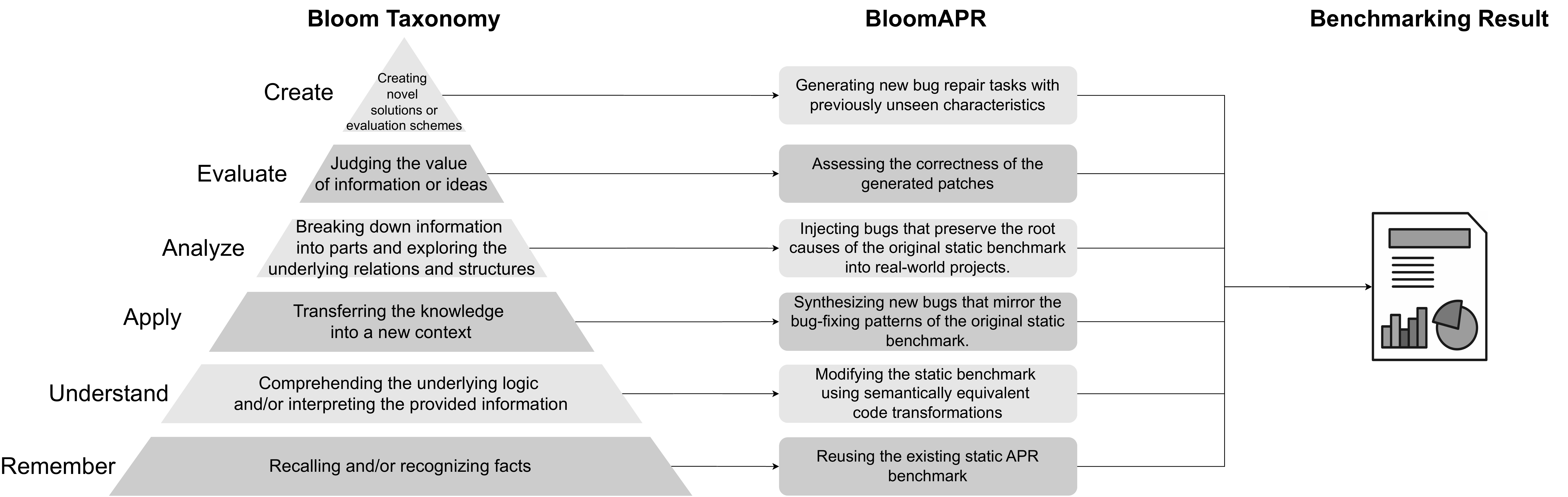}
    \caption{\normalsize Our \tool Framework.}
    \label{methodology}
\end{figure*}

\add{To address the above three challenges, we argue that rather than focusing solely on the difficult task of eliminating overlap between the benchmark and the LLM's training datasets, the research community should prioritize the development of innovative evaluation methodologies that can robustly assess LLM-powered SE systems across diverse usage contexts. Hence, in this paper, we propose a novel evaluation framework, {\tool}, which systematically assesses the capabilities of LLM-powered systems by incorporating Bloom's Taxonomy~\cite{bloom1956, AndersonKrathwohl2001}. As shown in Figure~\ref{methodology}, Bloom's Taxonomy classifies cognitive abilities into six levels of increasing complexity, ranging from lower-order thinking skills that require basic cognitive processing (namely \texttt{Remember}, \texttt{Understand}, and \texttt{Apply}) to higher-order thinking skills that require more advanced and deeper cognitive processing (namely \texttt{Analyze}, \texttt{Evaluate}, and \texttt{Create}). As our current prototype focuses on benchmark adaptations that can be instantiated and evaluated concretely, this paper operationalizes the first four layers (\texttt{Remember}, \texttt{Understand}, \texttt{Apply}, and \texttt{Analyze}).}

\add{Originally designed for educational settings, Bloom's Taxonomy has been widely adopted to structure learning objectives and design assessments. Integrating Bloom's Taxonomy into our evaluation framework enables fine-grained identification of specific competencies demonstrated by a solution and highlights areas for improvement~\cite{zhang2025UnseenHorizons, huber-niklaus-2025-llms}. Since our framework builds upon existing static benchmarks, the evaluation results are easily trackable and comparable, thereby ensuring consistency. Furthermore, our framework naturally aligns with the capabilities required for a mature and production-ready APR solution, including recognizing bug contexts and patterns, identifying root causes, and adapting repair strategies to different project contexts. To demonstrate the feasibility and effectiveness of our framework, we conducted case studies across two well-known static APR benchmarks (\defectj~\cite{Defects4J} and \SWE~\cite{jimenez2024swebench}), evaluating three representative LLM-powered APR systems (ChatRepair~\cite{10.1145/3650212.3680323}, CigaR~\cite{hidvégi2024cigarcostefficientprogramrepair}, and \SWEA~\cite{yang2024sweagent}).} 

\add{For evaluation, we report three evaluation metrics throughout the study: \emph{Plausible Patch (PP)}, \emph{SYntactic Equivalence (SYE)}, and \emph{Exact Match (EM)}. Our \defectj evaluation results show that these systems remain relatively strong on the memory-permissive \texttt{Remember} layer, with PP ranging from 53.92\% to 82.95\%. Under the \texttt{Understand} layer, where function and variable names are perturbed through \texttt{Hash-based} and \texttt{Rephrasing-based renaming} while preserving the original bug logic, PP drops to 21.66\%-57.60\% under \texttt{Hash-based renaming} and 28.11\%-55.76\% under \texttt{Rephrasing-based renaming}, showing that current systems remain sensitive to weakened lexical cues. In contrast, the \texttt{Apply} layer yields much stronger results on the synthetic bug-pattern dataset, where bug-fixing knowledge is transferred into newly reconstructed but still lightweight local contexts, with PP ranging from 72.35\% to 94.47\%. Finally, the \texttt{Analyze} layer remains the most challenging setting: when the same underlying root causes must be repaired in substantially different real-world project environments, PP ranges only from 2.88\% to 23.08\%, while SYE and EM both range from 0.00\% to 10.58\%. A secondary case study on \SWE shows a broadly similar qualitative contrast: lexical perturbations still reduce performance at \texttt{Understand}, synthetic \texttt{Apply} cases remain easier than the real-world \texttt{Analyze} setting, and the quantitative values vary with the benchmark and the amount of repository-level context available to the APR system. These benchmark-specific details are discussed later in Sections~\ref{sec:case} and~\ref{sec:swebench}. These results highlight a clear gap between benchmark-level success, controlled synthetic transfer, and robust real-world contextual transfer.} The contributions of this paper are: 

\begin{itemize}[leftmargin=*]
    \item This paper introduces \tool, the first systematic framework that transforms an existing static APR benchmark into a structured, dynamic evaluation environment grounded in Bloom's Taxonomy. Unlike existing benchmarking approaches, \tool organizes evaluation tasks into layers that diagnose different cognitive capabilities of LLM-powered APR systems. Our framework mitigates benchmark-contamination risk and enables systematic assessment of robustness, consistency, and patch faithfulness. To our knowledge, this is the first work to transform existing static benchmarks into structured dynamic evaluations grounded in Bloom's Taxonomy.

    \item \add{While we reuse existing memorization checks where applicable at the \texttt{Remember} layer, we introduce controlled lexical perturbations, synthetic bug-pattern generation, and root-cause-preserving real-world bug injections to instantiate the \texttt{Understand}, \texttt{Apply}, and \texttt{Analyze} layers, respectively. This layered structure enables not only raw accuracy reporting but also cross-layer insights into why APR systems succeed or fail under progressively more demanding bug-fixing knowledge transfer settings.}

    
    \item Prior empirical studies on the impact of data contamination in software engineering tasks (e.g.,~\cite{ramos2025largelanguagemodelsmemorizing, zhang2025UnseenHorizons}) have focused on direct LLM prompting. In contrast, \tool evaluates APR systems that employ agentic workflows (reflection, test-time sampling, iterative debugging), which iteratively refine bug fixes using the compiler, testing results, and contextual cues (e.g., code snippets or test cases). Our experiments reveal that these workflows do not automatically mitigate contamination effects, highlighting new research needs in reasoning and generalization for APR systems. To our knowledge, this is the first work to systematically evaluate the effectiveness of agentic mechanisms under contamination in a controlled setting.

    \item \tool further supports extensibility, allowing the incorporation of new benchmarks (e.g., \textsc{SWE-bench}~\cite{jimenez2024swebench}) and additional evaluation scenarios (e.g., code comment injection). In addition to performance benchmarking, \tool can also be useful for debugging APR systems and tuning LLM pipelines. 
    
    \item \add{We showed that existing static bug-fixing SE benchmarks (e.g., \defectj and \SWE) do not robustly or faithfully assess the capabilities of APR systems, as performance can differ sharply across memory-permissive, lexical-robustness, controlled-transfer, and real-world root-cause-transfer settings. These findings highlight the urgent need to evolve static benchmarks into more reliable dynamic evaluations.}

    \item To facilitate replication and further research, we shared our framework and evaluation results at~\cite{4openAnonymousGithub}.
\end{itemize}

\noindent \textbf{Paper Organization.} The rest of this paper is organized as follows: Section~\ref{sec:background} presents the background and reviews prior work related to this area. Section~\ref{sec:framework} describes our methodology for transforming existing static benchmarks into dynamic Bloom's Taxonomy-based benchmarks. Sections~\ref {sec:case} and~\ref{sec:swebench} present the results of our \defectj and \SWE case studies. Section~\ref{sec:discussion} discusses the findings and their implications. Section~\ref{sec:threats} outlines the threats to validity, and Section~\ref{sec:conclusion} concludes the paper.

\section{Background and Related Work}
\label{sec:background}


In this section, we discuss the background and existing research related to our work. 

\subsection{Automated Program Repair (APR)}
\label{sec:related_apr}
APR solutions automatically identify and fix software bugs without human intervention. 
Traditional APR solutions often rely on generate-and-validate approaches, where candidate patches are generated and tested against a suite of test cases to ensure correctness~\cite{10.1145/3293882.3330577, capgen, yuan2018arja, jiang2018shaping}.
Learning-based APR techniques utilize various information sources (e.g., historical bug-fix data) to predict potential patches, thereby improving the efficiency and accuracy of repairs~\cite{cure, sequencer, selfapr, rewardrepair}. 
Large Language Model (LLM)-based APR solutions utilize LLMs' advanced capabilities to identify and fix bugs in software code automatically.
Unlike traditional generate-and-validate approaches, LLM-powered APR solutions leverage the extensive pre-trained knowledge of LLMs to generate potential patches directly from the context of the entire program~\cite{bouzenia2024repairagent, yangmorepair, ahmed2023better, li2025hybrid}. Consistent with prior work~\cite{llmaprstudy, 10.1145/3650212.3680323, hidvégi2024cigarcostefficientprogramrepair, alpharepair}, we evaluated the effectiveness of APR systems using well-known metrics like Exact Match (EM), SYntactic Equivalence (SYE), and Plausible Patches (PP).

\subsection{Data Contamination}
\label{sec:related_data_conta}
Data contamination occurs when information from the test set is unintentionally included in the training set, leading to inflated performance metrics and compromised model evaluation.
In the context of LLMs, data contamination can significantly affect evaluation fairness, making it difficult to assess the model's true capabilities accurately.
Prior studies~\cite{renzullo2025automated, 10.1145/2635868.2635929, JiangICSE2023, silva2024repairbenchleaderboardfrontiermodels} have witnessed such influence and proposed various ways to mitigate the problem.
To avoid data contamination, Jiang et al. constructed HumanEval-Java~\cite{JiangICSE2023} by manually converting HumanEval Python programs to Java and injecting bugs into the correct Java programs.
\add{Meanwhile, RepairBench~\cite{silva2024repairbenchleaderboardfrontiermodels} was proposed as a standardized framework for evaluating frontier APR systems. It includes GitBug-Java~\cite{gitbugjava}, which contains bugs collected from 2023, which is beyond the training cut-off date of most LLMs—and \defectj. Because these datasets remain static, however, contamination risks may re-emerge as models are retrained on newer data~\cite{xu2024benchmarking, roberts2023data, deng2024investigating}. }

To assess the impact of data contamination and the ability of APR systems to generalize, we not only reused some existing approaches but also introduced novel evaluation techniques. In particular, we reused existing membership inference techniques~\cite{zhang_2025,274574} at the \texttt{Remember} layer to examine the overlap of the training dataset and the target benchmark. However, this result alone does not reveal the impact of memorization on APR effectiveness. Therefore, we re-ran and re-evaluated APR systems on the original benchmark and evaluated the results to assess their impact. The name perturbation technique at the \texttt{Understand} layer is similar to~\cite{zhang2025UnseenHorizons}. However, as we want to assess the effectiveness of APR under semantic variations, we evaluated the impact of two perturbation techniques: \texttt{Hash-based renaming} and \texttt{Rephrasing-based renaming}. We then introduce synthetic bug-pattern generation at the \texttt{Apply} layer and root-cause-preserving real-world bug injection at the \texttt{Analyze} layer to assess controlled transfer and real-world contextual transfer, respectively. 

\subsection{Dynamic Benchmarking}
\label{sec:related_dynamic_bench}
Dynamic benchmarking has emerged as a promising approach for evaluating LLMs, mitigating the limitations of static benchmarks that are prone to data contamination and fail to keep pace with rapidly advancing generation capabilities~\cite{wang2025benchmark, castillobeyond, kurtic2024mathador}.
Key features of dynamic benchmarking include: (1) adaptive evaluation, tailoring more complex tasks as models advance over time~\cite{zhudyval, zhangdarg}; (2) mitigation of data contamination by generating novel, out-of-distribution scenarios~\cite{chen2025dynamic,huang2025thinkbench}; and (3) fine-grained capability assessment that dissects specific cognitive abilities such as reasoning and problem-solving~\cite{zhu2024dyval, huang2025thinkbench}.
For instance, DyVal~\cite{zhudyval} utilizes directed acyclic graphs (DAGs) to create structured reasoning tasks with controllable complexity, while DyVal2 employs meta-probing agents inspired by psychometric theories to transform existing problems into new challenges targeting specific cognitive skills.
Similarly, DARG~\cite{zhangdarg} constructs and perturbs reasoning graphs to generate diverse test samples, enabling a more comprehensive evaluation of LLMs' generalization and robustness. 

While dynamic benchmarking has gained traction in evaluating LLMs, it remains underexplored in APR tasks.
Recent studies have introduced dynamic benchmarking frameworks in ASE that apply semantic-preserving mutations to code benchmarks, creating syntactically new yet semantically identical programs~\cite{chen2025dynamic, guan2025your, orvalho2025large}.
These approaches aim to mitigate data contamination and better assess the robustness of code understanding and reasoning models.
However, the integration of such dynamic benchmarking methodologies for the APR task remains unexplored.
To address these challenges, we propose the \tool framework, which incorporates dynamic benchmarking to robustly and reliably evaluate LLMs for the APR task by adapting to their evolving capabilities at various cognitive layers and mitigating data contamination.
\section{Our Evaluation Framework}
\label{sec:framework}

In this section, we describe our evaluation framework, {\tool}, which systematically assesses the capabilities of LLM-powered APR solutions. As illustrated in Figure~\ref{methodology}, our framework is conceptually divided into six layers, corresponding to Bloom's Taxonomy: \texttt{Remember}, \texttt{Understand}, \texttt{Apply}, \texttt{Analyze}, \texttt{Evaluate}, and \texttt{Create}. \add{The current version of \tool operationalizes only the first four layers. We do not claim a one-to-one equivalence between human cognition and LLM behavior. Instead, we use these four layers as an evaluation ladder that progressively assesses the capabilities (recognizing bug contexts and patterns, identifying root causes, and adapting repair strategies to different project contexts) of robust and production-ready APR systems in increasing levels of complexity. \respto{E0-0.1} \respto{E0-3.4} \respto{R1-0.1} \respto{R1-1.1} \respto{R1-14.1}} 


Our framework leverages high-quality data from publicly available static bug-fixing benchmarks to facilitate cross-comparison and performance tuning. Depending on the objectives of each layer, variations of each bug instance from the adopted benchmarks are generated using techniques such as program transformation, data synthesis, and bug injection. Bloom's taxonomy is hierarchical, meaning each layer builds on and assumes the competencies of those below it. To demonstrate the feasibility and effectiveness of \tool, we developed our research prototype based on version $2.1.0$ of the \defectj benchmark. \add{In our design, the \texttt{Remember} layer evaluates whether LLM-powered APR systems under evaluation (SUEs) can retrieve or recognize known fixes or patterns from a bug-fixing benchmark. For each bug instance in the original \defectj, the subsequent layers deliberately increase difficulty and novelty in a controlled manner: \respto{R1-0.2}} 

\begin{itemize}
    \item \emph{Lexical variations}: the \texttt{Understand} layer evaluates whether SUEs can still comprehend local code semantics when human-readable identifier cues are perturbed through behavior-preserving renaming.
    \item \emph{Controlled semantic transfer}: the \texttt{Apply} layer evaluates whether SUEs can transfer repair knowledge to newly instantiated synthetic bugs in a simple but different project context that preserves the bug pattern.
    \item \emph{Context variations}: the \texttt{Analyze} layer further increases complexity by injecting synthetically generated bugs with the same root causes into diverse real-world project contexts with external dependencies. These synthetically generated bugs may exhibit little or no textual similarity to the original bugs. 
\end{itemize}

\add{In this initial prototype, we did not address the top two advanced reasoning layers in {\tool}: \texttt{Evaluate} and \texttt{Create}. The \texttt{Evaluate} layer focuses on patch assessment and selection, which is often treated as a related but distinct research problem (e.g.,~\cite{zhou2023patchzero,le2023invalidator,le2019reliability}). The \texttt{Create} layer concerns the reliable generation of genuinely novel repair tasks, bug types, and corresponding fixes. This capability is not operationalized or evaluated in the current BloomAPR prototype. 
The remainder of this section explains each layer in detail. 
To aid explanation, we use a running example to illustrate how an existing bug instance from \defectj is transformed across different layers in Figure~\ref{fig:jc5}.}

\add{Importantly, these layers should be understood as evaluation objectives rather than as one-to-one mappings to fixed mutation operators. In the current research prototype, we instantiate each operationalized layer through one primary transformation family to facilitate illustration and keep the two case studies (\defectj and \SWE) controlled, comparable, and easy to interpret. This design choice does not imply that each layer is exhausted by a single operator. Rather, it means that the current prototype provides one concrete realization of a broader layered methodology. 
We return to the broader realizations and their supplementary validations in Section~\ref{sec:discussion_layer_generalization}.
 \respto{E0-2.2}}

\subsection{Remember}

\noindent \textbf{Objective}: \add{The goal of the evaluation at the \texttt{Remember} layer is to assess the performance of the LLM-powered APR systems under evaluation (SUEs) on the original bug-fixing benchmark (\defectj), where benchmark familiarity and memorized recognition from pretraining datasets may contribute.}

\noindent \textbf{Approach}: \add{Prior work suggests that widely used bug benchmarks may already overlap with LLM training data~\cite{ramos2025largelanguagemodelsmemorizing, zhang2025UnseenHorizons}. 
We therefore use the original \defectj setting as a memory-permissive reference baseline, because it preserves the benchmark in its original form and maximizes the availability of benchmark-specific lexical cues and possible prior exposure. We do not claim that SUE performance at this layer reflects pure memorization with zero reasoning. Instead, this layer measures performance in a setting where memorized recognition and benchmark familiarity from training datasets may substantially contribute.} 


\noindent \add{Hence, at the \texttt{Remember} layer of \tool, we selected 217 bug instances (single-function-level bugs) in \defectj v2.1.0, and intentionally excluded bugs from the earlier v1.0.0 release. We applied each APR system under evaluation (SUE) to these 217 bug instances to establish a reference performance level under the original benchmark setting, where previously observed bug-fixing patterns or benchmark exposure can be captured and made available to the model as a result of its training. We use this layer as the baseline because it preserves the benchmark in its original static form and is therefore the closest to the evaluation setting commonly used in prior APR studies. The subsequent layers then introduce progressively stronger perturbations, synthetic reconstructions, or real-world contextual variation relative to this reference point, allowing us to interpret how performance changes as the evaluation moves beyond the original benchmark conditions.}


\noindent \textbf{Evaluation Schemes}: These three evaluation metrics will be tracked: 

\begin{itemize}[leftmargin=*]
    \item \% bugs with \emph{plausible patches (PP)}: PPs refer to bug fixes that pass all the given tests. \% of bugs with PP is an optimistic measure, as some of the plausible patches may fail on additional test cases that are not included. Hence, it provides an upper bound in terms of how well the evaluated solutions can recognize known bug-fixing patterns.

    \item \add{\% bugs with \emph{SYntactic Equivalence (SYE)} patches: 
    SYE refers to a plausible patch whose normalized token sequence matches that of the developer patch after comments and whitespace are removed and variable identifiers are consistently replaced with generic placeholders. SYE captures normalized structural similarity to the developer patch; it does not independently establish semantic or behavioral equivalence. For each PP and its corresponding oracle code snippets, we removed comments and whitespace, then tokenized the resulting code snippets and replaced variable names with generic placeholders. If the resulting sequences of tokens were identical, the patch was classified as an SYE. \respto{R1-7.1}}
    
    \item \% bugs with \emph{exact matches (EM)} patches:
    EMs refer to bug fixes that exactly match the provided oracle after comments and whitespace are removed. While there may be multiple valid ways to fix a bug, a high percentage of EMs is a strong indicator of the SUEs' effectiveness. For each PP and its corresponding oracle code snippets, we removed comments and whitespace, then compared the resulting string sequences directly. If they were equal, the patch was classified as an EM. 
\end{itemize}

\begin{figure*}[htbp]
    \centering
    \includegraphics[width=1.00\textwidth]{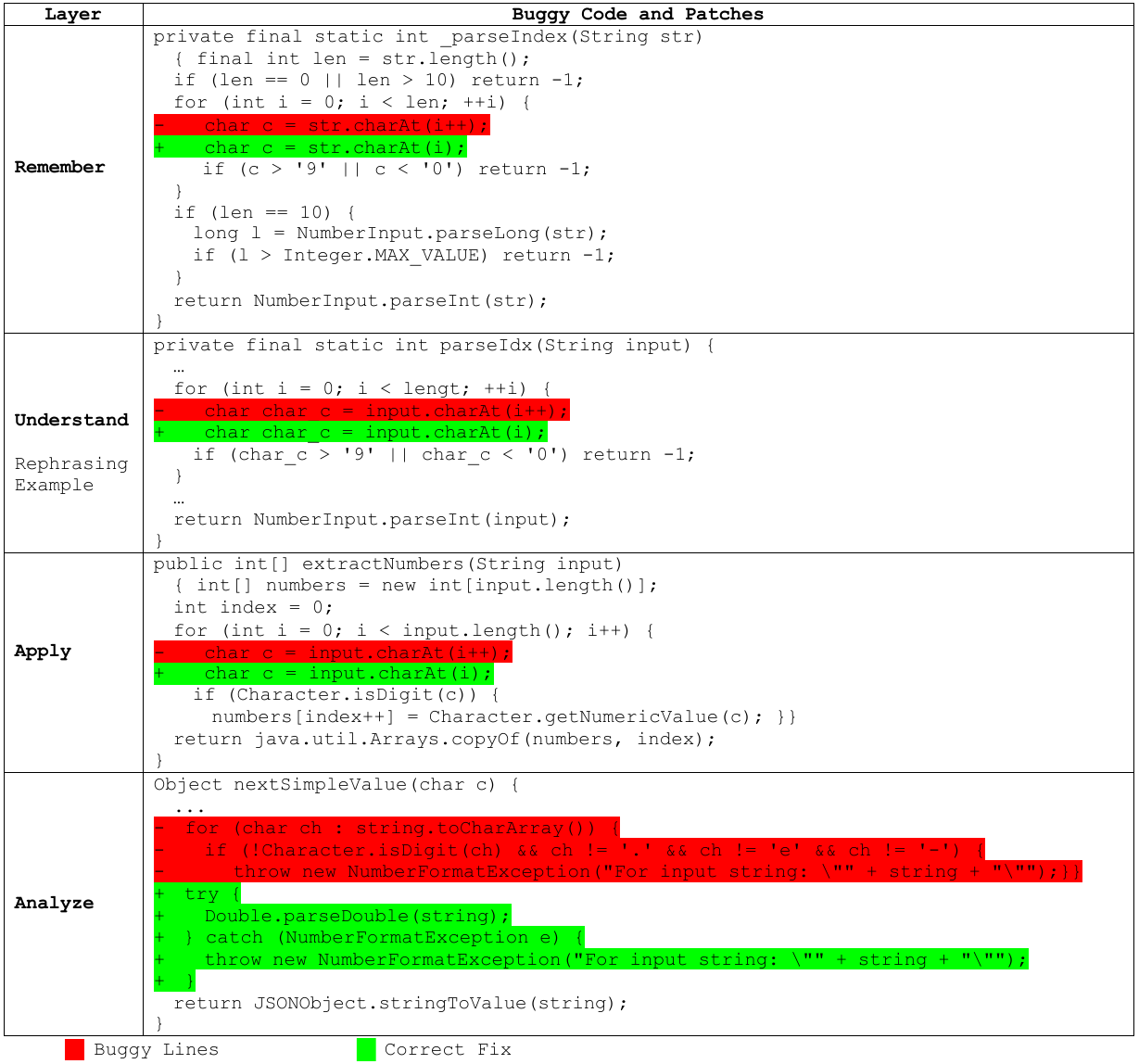}
    \caption{\normalsize 
    \add{Code snippets for the \texttt{JacksonCore-5} bug adapted across the four operationalized layers of \tool. All variants preserve a semantic relationship to the original defect, while four layers introduce different sources of difficulties. All these bug variants share the same root cause (i.e., they incorrectly process numeric-string inputs during character traversal or validation) and can be triggered by the same test case (``\texttt{1e0}''). The project/filename is \texttt{JacksonCore/JsonPointer.java} for the \texttt{Remember}, \texttt{Understand}, and \texttt{Apply} layer examples, and \texttt{JSON-java/JSONTokener.java} for the \texttt{Analyze} layer example.}}
    \label{fig:jc5}
\end{figure*}

\begin{figure*}[htbp]
    \centering
    \includegraphics[width=1.05\textwidth]{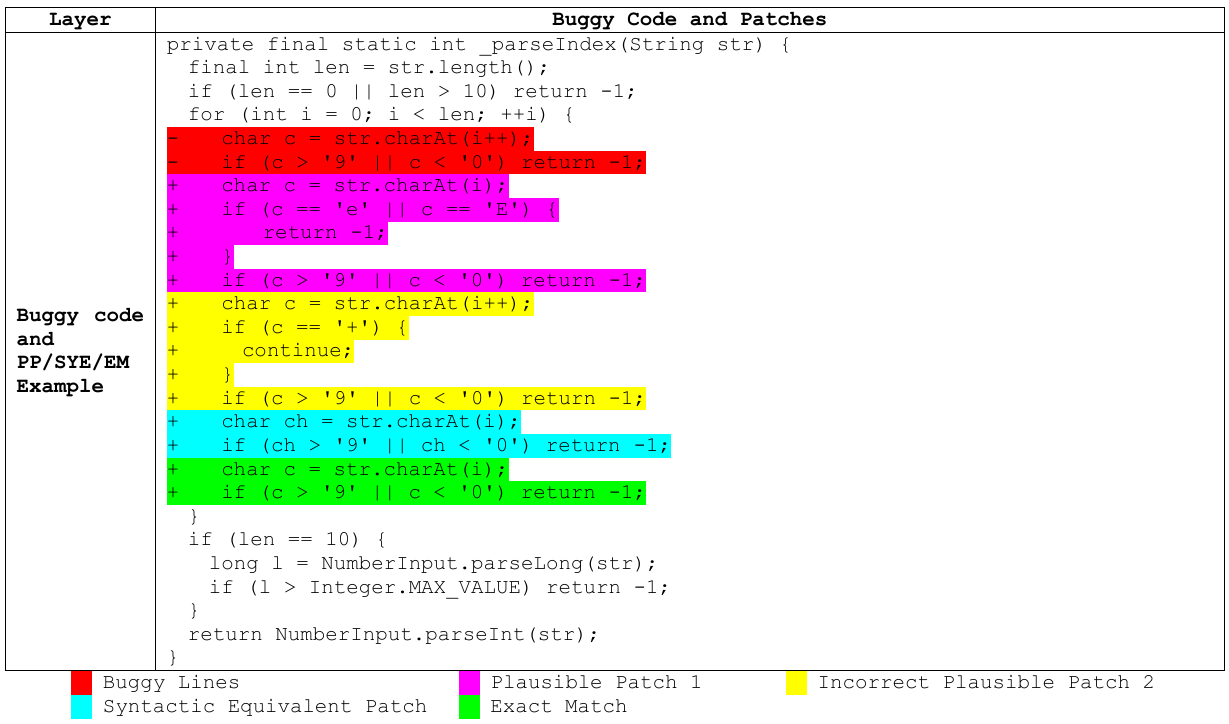}
    \caption{\normalsize PP/SYE/EM patch examples from the \texttt{JacksonCore-5} bug.}
    \label{fig:jc5_case}
\end{figure*}

Figure~\ref{fig:jc5} shows an example bug instance (\texttt{JacksonCore-5}) from the \defectj benchmark. It is a bug in a parsing function that converts a string input into a number. The code block highlighted in red indicates the buggy code, while the block highlighted in green shows the fix. This bug stems from a failure to correctly handle scientific notation. The test case (``\texttt{1e0}''), which represents the string form of a number in scientific notation, triggers the failure. At the \texttt{Remember} layer, we retained the original \texttt{JacksonCore-5} bug, its associated bug fix, and test cases without modification. Figure~\ref{fig:jc5_case} shows an example of PP/SYE/EM patches for the same \texttt{JacksonCore-5} bug. Both PPs pass the existing test cases. However, \texttt{Plausible Patch 2} is invalid, as the code block fails to deal with some invalid inputs (e.g., \texttt{\_}, \texttt{+}, or blank space), which are not considered in the current test cases.

\subsection{Understand}
\noindent \textbf{Objective}: \add{The goal of the evaluation at the \texttt{Understand} layer is to determine whether the SUEs can still understand the logic of the code and apply appropriate bug-fixing patterns when familiar bugs are perturbed through lexical variations. \respto{E0-0.2}}


\noindent \textbf{Approach}: \add{Various behavior-preserving code transformations can be applied here, so that the textual context of the code is altered while keeping the runtime behavior the same. As shown in previous studies, LLM-powered APR systems can be brittle even in response to small code perturbations~\cite{WuISSTA2023,zhang2025UnseenHorizons,wang2022recoderobustnessevaluationcode}. In this layer, we apply simple code transformation actions (variable and function renaming) as controlled perturbations to examine whether SUEs can still understand the code's logic when surface lexical cues are altered. Altering \emph{variable names} is intended to test whether the SUEs can follow the internal logical flow of the code within a function, while altering \emph{function names} is intended to test whether the SUEs can still interpret function-level dependencies when descriptive naming cues are changed. These two types of renaming schemes were applied:\respto{R1-0.3} \respto{R1-1.2}}


\begin{enumerate}
    \item \emph{Hash-based renaming}: Instead of texts that can be read and parsed for meaning, renaming function/variable names with cryptic hash values forces the SUEs to comprehend the underlying logic behind the code, as these names are no longer meaningful and cannot provide any clues regarding the purpose or the concrete implementation of the buggy code. \add{This tests whether the SUEs can still resolve the bug by relying primarily on code structure and semantics rather than meaningful identifier names.}
    \item \emph{Rephrasing-based renaming}: 
    Depending on the context of the code, alternative function and variable names are given. The rephrased variable/function names can be misleading or contextually different from the original names. Hence, the SUEs must understand the bug in context and reconcile potentially tricky naming cues to fix the bugs. 

\end{enumerate}

We first used the Eclipse AST package~\cite{eclipse_ast_package} to locate which buggy functions and their internal variables should be renamed. Then, to generate names for the Rephrasing-based renaming scheme, we fed the buggy functions to GPT-4o and prompted the model to provide alternative function names and all the internal variables within as outputs. To generate names under the \texttt{Hash-based renaming} scheme, we wrote scripts to calculate the SHA-256 hashes for the target function names and their internal variable names, respectively. To ensure the uniqueness of the generated hash values, we concatenated these names along with the current timestamp before calculating their hash values. 

Finally, we replaced these names in the original buggy functions with the generated alternative names by again parsing through the ASTs of these buggy functions. For the renamed functions, we performed an additional pass to ensure all the invocations of these functions from the feature code and the test code in this project are updated. 

For our research prototype, we applied the transformations to the same 217 \defectj v2.1.0 bug instances. Among these cases, we managed to rename all the internal variable names of the buggy functions. We only managed to rename function names for 183 bug instances due to inheritance and function overloading for the other 34 instances. 
The \texttt{Understand} row in Figure~\ref{fig:jc5} shows an example of the rephrased version of the bug. The function name \texttt{\_parseIndex} was renamed to \texttt{parseIdx}, and all internal variables were renamed accordingly. The call chain within the project was also updated to ensure successful compilation. Since the test suite does not directly invoke this function, no modifications were required for the test cases. Due to space constraints, please refer to our replication package for an example of \texttt{Hash-based renaming}.

\noindent \textbf{Evaluation Schemes}: Similar to the \texttt{Remember} layer, we tracked the \% of bugs with PPs/SYEs/EMs. 

\subsection{Apply}
\noindent \textbf{Objective}: \add{The goal of the evaluation at the \texttt{Apply} layer is to assess whether SUEs can apply prior bug-fixing knowledge to textually altered synthetic bugs. These synthetic instances retain the core buggy patterns of the original benchmark, but are presented within newly constructed local contexts.\respto{E0-0.3} \respto{R1-0.4}}

\noindent \textbf{Approach}: We leveraged LLMs to first analyze each bug instance from the original static benchmark and then to generate novel synthetic bugs with corresponding test cases. These synthetic cases were created based on the LLM's analysis of each bug by preserving similar buggy behavior patterns from the original bug instances.

\add{For our prototype, we selected a different LLM, Claude 3.5 Sonnet, to analyze the benchmark bugs and generate the synthetic datasets. We chose Claude 3.5 Sonnet primarily for its generation quality and because it is outside the set of LLMs later evaluated in our case study. This choice does not guarantee the absence of data contamination in the strict sense. Rather, it is intended to reduce benchmark-construction bias by mitigating two risks: \emph{Shared Canonical Representation (Token Consistency)}~\cite{yuan2026silencer}, where the generator and the evaluated model may benefit from similar stylistic or token-level preferences, and the \emph{homework problem} (i.e., self-correction bias)~\cite{panickssery2024llm}, where a model is later evaluated on artifacts that it or a closely related model family effectively helped author.\respto{R1-8.1}
Under this setup, we generated one primary synthetic bug-pattern dataset group:}

\begin{itemize}[leftmargin=*]
    \item \emph{Synthetic Bugs with Similar Buggy Behavior}: \add{For each bug instance in the original static benchmark, we prompted Claude to generate one set of buggy code, bug fixes, and corresponding test cases using the buggy functions, fixed code snippets, and error messages of the original bug as input. The goal of this variant is to preserve a similar observable buggy behavior pattern while placing it in a new synthetic local context.}

\end{itemize}

\add{The \texttt{Apply} row in Figure~\ref{fig:jc5} illustrates the role of our synthetic bug-pattern dataset. In this setting, each generated bug is designed to exhibit a failure pattern similar to the original benchmark instance, such as producing an incorrect parsing result or triggering a comparable assertion failure, while appearing in a new synthetic local context. Although the surrounding code structure, identifier usage, and implementation details may differ substantially from the original benchmark instance, the synthetic case preserves a similar buggy behavior pattern and therefore allows us to evaluate whether SUEs can transfer bug-fixing knowledge beyond the original benchmark code.\respto{R1-1.3}}

\add{This type of evaluation differs from the \texttt{Understand} layer because the perturbation here is not limited to lexical renaming within the original benchmark instances. Instead, the bugs, fixes, and tests are reconstructed into entirely new synthetic local contexts by an LLM.}

\noindent \textbf{Evaluation Schemes}: Similar to the \texttt{Understand} layer, we tracked the \% of bugs with PPs/SYEs/EMs.

\subsection{Analyze}

\noindent \textbf{Objective}: \add{The goal of the evaluation at the \texttt{Analyze} layer is to assess whether SUEs can identify and repair bugs that preserve the same underlying root cause as the benchmark instances, while appearing in substantially different real-world project contexts with different build systems, dependencies, and usage scenarios.\respto{E0-0.4}}


\noindent \textbf{Approach}: 
\add{At the \texttt{Apply} layer, we evaluated SUEs on synthetic bugs derived from the original benchmark while preserving similar buggy behavior in new local contexts. Here, we evaluate whether SUEs can identify and repair bugs that preserve the same underlying root cause as the benchmark instances, but appear in substantially different real-world project environments.}
\add{This is more challenging because, beyond understanding the structure and logic of the corresponding code snippets, the SUEs must also break down unfamiliar project contexts (e.g., external dependencies, deployment environment, and usage scenarios) into relevant and irrelevant parts, and then determine which local operations, state transitions, or missing checks correspond to the same latent failure root cause as in the benchmark bug. To ensure the freshness of our evaluation datasets, we injected new bugs with the same underlying root causes as the benchmark instances into real-world projects. Importantly, the goal of this layer is not to reproduce the same buggy pattern or patch shape from the original benchmark, but to preserve the same failure mechanism in a substantially different real-world environment. In this sense, the increase in cognitive complexity comes from requiring the SUEs to abstract away from surface-level similarity and recover the same deeper failure mechanism across substantially different contextual realizations. Our steps are listed below:\respto{R1-2.2} \respto{R1-0.5}}

\begin{itemize}
    \item \emph{Step 1 - Code Search}: For each bug instance in the target benchmark, we searched for open-source projects containing code snippets similar to the fixed benchmark versions. In other words, we wanted to first search for the correct version of the code from the real-world project and then inject bugs afterward. We leveraged a code search engine to identify real-world code snippets that are semantically related to the fixed benchmark instances, as such tools can jointly reason over natural language and code within a broader context. The search results contain not only code snippets with exact textual matches (a.k.a., type-1 clones) but also functions with small variations (a.k.a., type-2 or type-3 clones). \add{Unfortunately, GitHub Copilot Chat~\cite{microsoft_copilot}, which we used in March 2025 as an LLM-assisted code search tool for constructing the \defectj-based real-world \texttt{Analyze} dataset, sometimes hallucinated by returning phantom projects or non-existent code snippets.}
    \add{Since some bug instances do not have any matches in real-world projects (e.g., \texttt{Cli-5}), after filtering and validation, the metadata pool contains 261 real-world cases spanning 37 project names. For the final one-to-one \texttt{Analyze} evaluation, we group metadata entries by each original \texttt{project\_bug} ID and select the first available case ID according to the numerical case-ID order for that group. This yields 104 matched Defects4J bug IDs, each represented by one validated real-world instance from 28 distinct real-world projects (noted as the 104-instance Analyze dataset). This selection did not use PP, SYE, EM, or any other SUE outcome.}

    \item \emph{Step 2 - Bug Injection}: We used LLMs to inject bugs through few-shot prompting. For our research prototype, all of the obtained code snippets from the previous step correspond to the fixed version of the selected bug instances. \add{We used GPT-4o to inject bugs into these code snippets by supplying the corresponding benchmark bugs and fixes as references, with the aim of reproducing the same underlying root cause in the new project context. As illustrated by the running example in Figure~\ref{fig:jc5}, the \texttt{Analyze} instance may look textually very different from the original benchmark case because it does not preserve the same buggy pattern (i.e., local patch shape). In the real-world \texttt{Analyze} example, the bug is re-instantiated in \texttt{nextSimpleValue(char c)}, where the code first constructs a token string and then validates whether it represents a legal numeric value. The buggy version performs this validation through manual character-by-character screening, explicitly iterating over \texttt{string.toCharArray()} and rejecting characters outside a restricted set. The repaired version replaces this ad hoc validation logic with a \texttt{try}-\texttt{catch} around \texttt{Double.parseDouble(string)}, throwing the same \texttt{NumberFormatException} only when parsing fails. Thus, what is preserved is not the visible edit form of the original benchmark patch, but the deeper failure root cause: incorrect handling of numeric-string parsing and validation under a substantially different real-world implementation context. \respto{R1-2.1}} 

    \item \emph{Step 3 - Test Case Generation}: Similar to the previous step, for each injected bug instance, we used LLMs to generate the corresponding test cases through few-shot prompting. We then reused the existing test harness, when available in the original project, and inserted the newly generated test cases into it. The one remaining project, BitStream~\cite{bitstream}, does not have any test cases. Therefore, we developed its test harness before injecting the newly generated tests.

    \item \emph{Step 4 - Verification and Refinement}:     
    The resulting buggy and bug-fixed versions were compiled and tested to ensure the validity of the bug/bug-fixing pairs and the effectiveness of the generated test cases. Whenever compilation errors or test execution issues arose (e.g., the test failing to reveal the bug or still failing on the fixed version), the error messages and the current implementations were fed back to GPT-4o for further refinement. This process was repeated until no further compilation or testing issues remained. \add{After the above process, we retained $104$ matched real-world cases that could be successfully injected, validated, and tested. We report SUE performance only on this finalized set. \respto{R1-2.3}}
\end{itemize}

\noindent \textbf{Evaluation Schemes}: \add{The \texttt{Analyze} results are reported on one finalized representative real-world instance per matched \defectj bug ID. Here, one-to-one means that each of the 104 matched \defectj bug IDs contributes exactly one validated real-world instance to the main \texttt{Analyze} evaluation; multiple Defects4J bug IDs may originate from the same real-world project. Similar to the \texttt{Understand} layer, we directly track the \% of bugs with PPs, SYEs, and EMs.}

\section{\defectj Case Study}
\label{sec:case}
\begin{figure*}[htbp]
  \centering
  \includegraphics[width=0.8\textwidth]{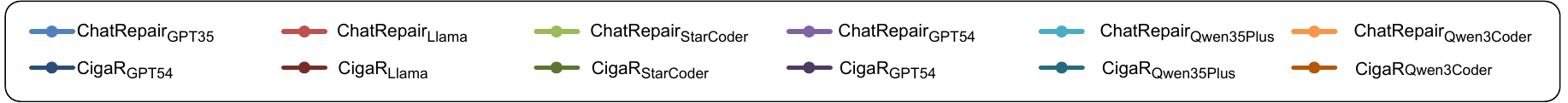}

    \begin{subfigure}[b]{0.97\textwidth}
        \includegraphics[width=\textwidth]{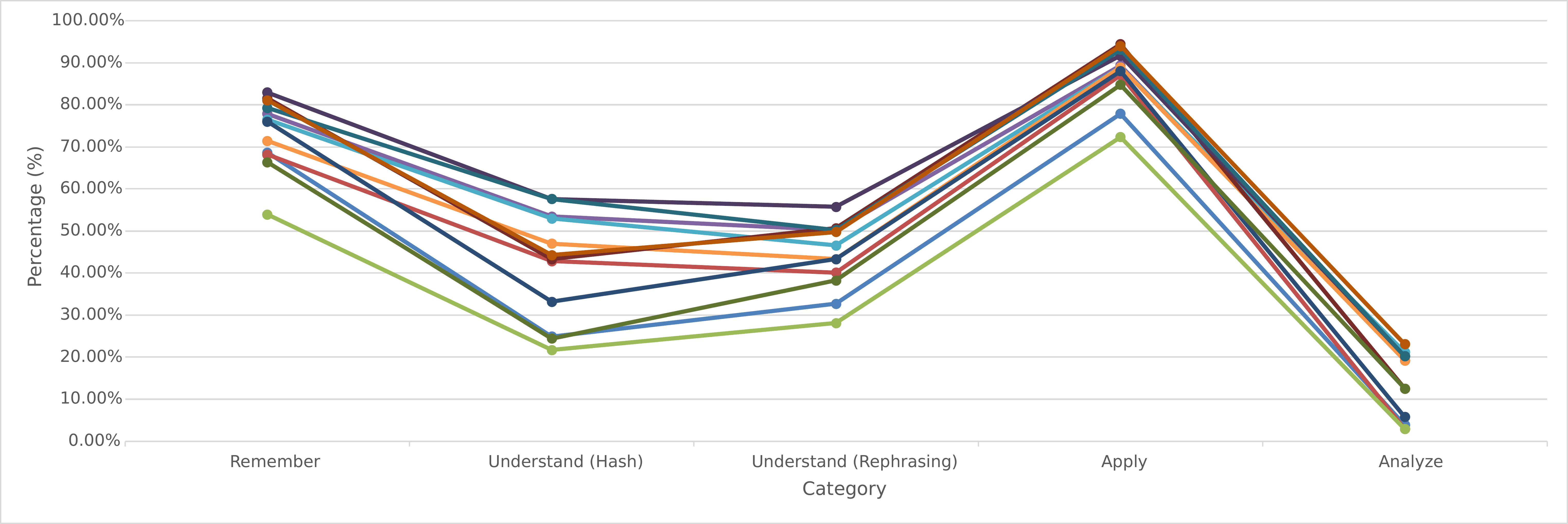}
        \caption{\footnotesize \% of Bugs with PP}
        \label{rq2_pp}
    \end{subfigure}
    \hfill
    \begin{subfigure}[b]{0.97\textwidth}
        \includegraphics[width=\textwidth]{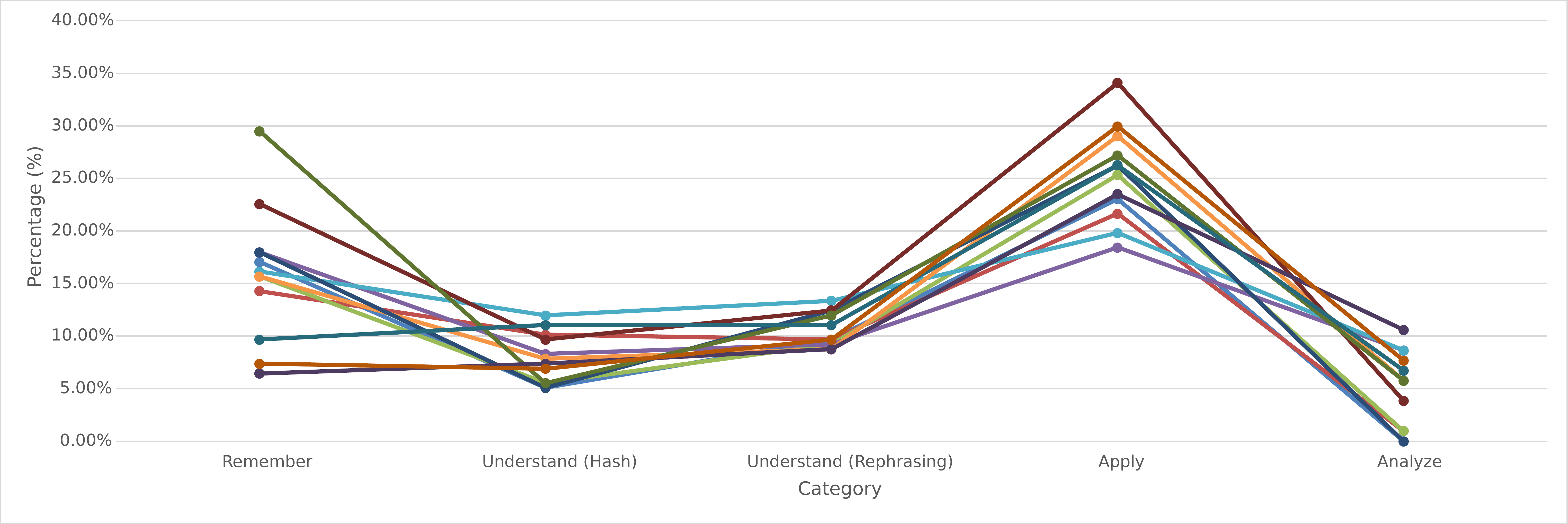}
        \caption{\footnotesize \% of Bugs with SYE}
        \label{rq2_sye}
    \end{subfigure}
    \hfill
    \begin{subfigure}[b]{0.97\textwidth}
        \includegraphics[width=\textwidth]{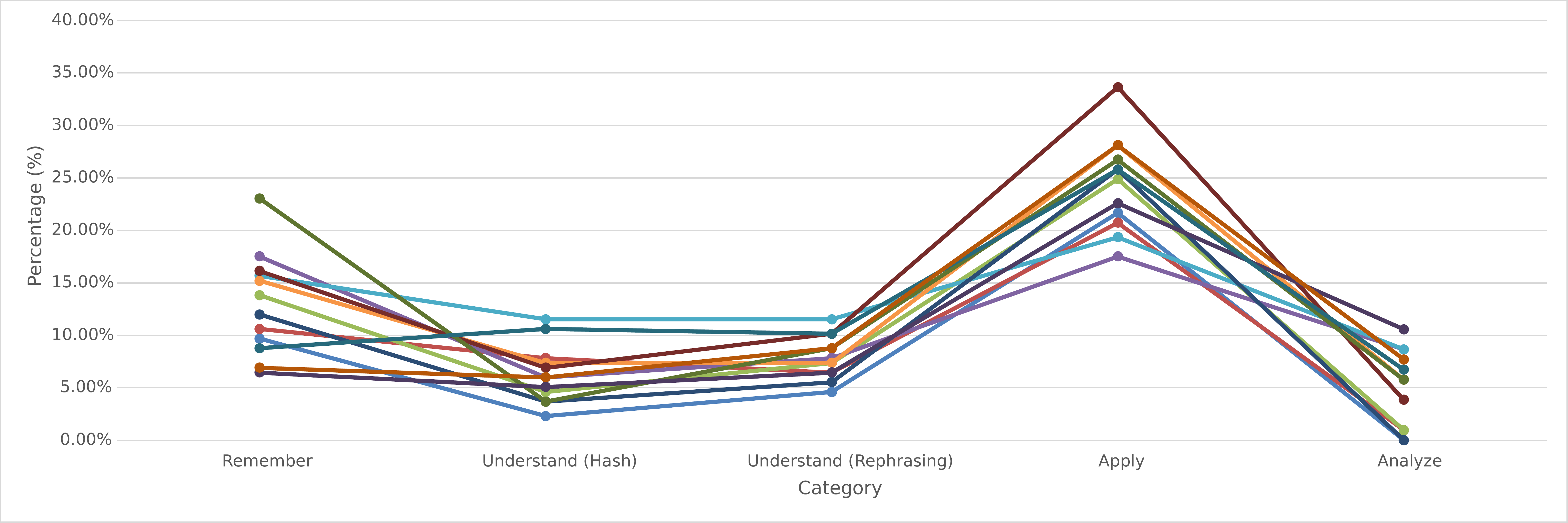}
        \caption{\footnotesize \% of Bugs with EM}
        \label{rq2_em}
    \end{subfigure}
    \caption{\add{PP, SYE, and EM Evaluation Results across different layers for \defectj. For the \texttt{Understand} layer, we report both \texttt{Hash-based renaming} (denoted as Hash) results and \texttt{Rephrasing-based renaming} (denoted as Rephrasing) results.}\respto{E0-3.2}\respto{R1-10.1}}
    \label{rq2}
\end{figure*}

\add{In the previous section, we introduced our framework, \tool, for systematically assessing the capabilities of LLM-powered APR solutions. In this section, we present our first case study, in which we apply \tool to the static \defectj benchmark and evaluate two representative LLM-powered APR systems: ChatRepair~\cite{10.1145/3650212.3680323} and CigaR~\cite{hidvégi2024cigarcostefficientprogramrepair}. 
Rather than directly prompting LLMs, both approaches invoke them through APR systems that iteratively generate and refine bug fixes using reflection, test-time sampling, and environmental feedback (e.g., compilation and testing results) or enriched problem context (e.g., failure data and examples).

To ensure consistency and generalizability, we selected six representative LLMs for evaluation based on their coding capability, availability, and deployment style: (i) GPT-3.5-turbo-1106~\cite{openai2025gpt35}, a closed-source model trained by OpenAI, which served as the default model in ChatRepair and CigaR, as per their replication packages; (ii) Llama3.1-70B, one of the most advanced open-weight LLMs trained by Meta~\cite{huggingface_llama_3_1_70B_instruct}; and (iii) StarCoder2-15B-Instruct-v0.1, an open-source model with publicly available training data and pipeline, specialized for code generation~\cite{wei2024selfcodealign}; (iv) GPT-5.4 mini~\cite{openai_gpt54mini}, a closed-source model optimized for coding and tool-based tasks; (v) Qwen3.5-Plus~\cite{qwen35plus}, a reasoning-oriented model with thinking capabilities enabled; and (vi) Qwen3-Coder~\cite{qwen3coder}, a coding-specialized model designed for agentic programming tasks. In the remainder of this paper, we refer to these models as GPT35, Llama, StarCoder, GPT54, Qwen35Plus, and Qwen3Coder for brevity. The last three models were selected to reflect the most recent model generation available at the time of the paper and were released from January 2026 onward. 
\respto{E0-1.1} \respto{R2-0.1} \respto{R2-1.2} \respto{R2-0.2}
Figure~\ref{rq2} summarizes the evaluation results for the twelve resulting setups (two SUEs $\times$ six LLMs) across the four layers of \tool under two renaming conditions. Detailed results for each layer are discussed in the remainder of this section. We use the notation of $SUE\nolinefrac{\scriptscriptstyle {Layer}}{\scriptscriptstyle {LLM}}$ to represent each configuration at a specific layer throughout the following discussion. \respto{R1-6.1} \respto{R2-1.3}
}

\subsection{Remember}
\label{sec:RQ1}
\add{As shown in Figure~\ref{rq2}, under the \texttt{Remember} layer on \defectj, the twelve evaluated SUE setups reported PP rates ranging from 53.92\% to 82.95\%, SYE rates ranging from 6.45\% to 29.49\%, and EM rates ranging from 6.45\% to 23.04\%. These results indicate that even in the original benchmark setting, plausible repair remains substantially easier than exact reproduction. We also observed clear variation across both APR systems and LLM backbones. For example, $CigaR^{_{Remember}}_{_{GPT54}}$ achieves the highest PP rate at 82.95\%, whereas $CigaR^{_{Remember}}_{_{StarCoder}}$ yields the highest SYE and EM rates at 29.49\% and 23.04\%, respectively. Across the six evaluated LLM backbones, the \texttt{Remember} results therefore remain sensitive to both the agent design and the underlying model configuration. Overall, the \texttt{Remember} layer should be interpreted as a memory-permissive baseline rather than as a setting in which any single model family uniformly dominates across all three metrics.}

\add{We conducted further examinations across all six evaluated model backbones to assess whether \defectj was likely included in their training data. For StarCoder2, we queried the released dataset utility~\cite{zhang_2025} and confirmed that \defectj projects were present in its training corpus. For the remaining models, whose training-data details were not publicly disclosed, we followed prior work~\cite{274574} and used prompt-based extraction to probe for benchmark exposure under a no-web-access setting. When queried about \defectj bugs such as \texttt{Compress-7}, the models were often able to recall the associated failure causes and reconstruct the corresponding fixes from highly limited cues. In particular, GPT35 and Llama correctly recalled the associated errors and completed the fixes for the corresponding code snippets, while GPT54, Qwen35Plus, and Qwen3Coder further reconstructed the full function body from only a partial function snippet and also recovered a relatively complete explanation of the bug cause without Internet access. In several cases, these newer models produced more complete reconstructions than the earlier ones. Taken together, these results provide strong evidence that prior \defectj exposure likely contributed to the \texttt{Remember}-layer performance across the evaluated model families.}

\add{Rather than interpreting the \texttt{Remember} layer as a proof of pure memorization, we treat it as a memory-permissive baseline in which prior benchmark exposure and familiar bug-fixing patterns are most available. This interpretation is consistent with the relatively high PP rates observed on \defectj and explains why exact reproduction remains limited: even in the original benchmark setting, EM remains substantially lower than PP across all twelve setups.}\respto{R1-0.6}\jack{whose comments you addressed here? Please mark}


\mybox{\textbf{Summary}: \add{On \defectj, the \texttt{Remember} layer yields relatively strong PP performance (53.92\%-82.95\%), but much lower SYE and EM rates (6.45\%-29.49\% and 6.45\%-23.04\%, respectively). This gap shows that even under the original static benchmark (i.e., the unaltered benchmark setting), generating a test-passing patch is substantially easier than reproducing the developer-style repair itself. In other words, the evaluated APR systems may still recall benchmark-related facts, familiar bug-fixing patterns, or partial repair cues, but their exact reproduction of the fixing details remains much more limited. These results therefore support our interpretation of the \texttt{Remember} layer as a memory-permissive reference baseline rather than as a direct proof of pure rote memorization.}}

\subsection{Understand}
\label{sec:RQ2}
\add{The two groups marked as ``Understand (Hash)'' and ``Understand (Rephrasing)'' \yinghang{Rephrasing, changed here and in the chart} in Figure~\ref{rq2} show the PP/SYE/EM results for the \texttt{Understand} layer under \emph{Hash-based renaming} and \emph{Rephrasing-based renaming}, respectively. In both settings, function and variable names were perturbed while preserving the original program behavior, so that the evaluation isolates whether the SUEs can still repair familiar bugs when human-readable identifier cues are weakened or removed. \respto{R1-4.1} 

Overall, we observed a clear degradation in PP performance compared with the \texttt{Remember} layer across all twelve setups. Under the combined function-and-variable renaming setting, PP drops from 53.92\%-82.95\%\yinghang{Jiho was right, this is old data.} at the \texttt{Remember} layer to 21.66\%-57.60\% under \texttt{Hash-based renaming} and to 28.11\%-55.76\% under \texttt{Rephrasing-based renaming}. This shows that even strong modern models remain sensitive to lexical perturbations when they can no longer rely on the original benchmark wording and identifier semantics. Compared with the \texttt{Remember} layer, these drops show that lexical perturbations alone are already sufficient to noticeably weaken repair performance, even before the additional transfer challenges introduced in the \texttt{Apply} and \texttt{Analyze} layers. These differences are also consistent with our statistical analysis: the PP reductions from the \texttt{Remember} layer to the combined-renaming \texttt{Understand} settings are significant under McNemar's test across all $24$ paired comparisons from two APR systems × six LLMs × two renaming conditions (\texttt{Hash-based renaming} and \texttt{Rephrasing-based renaming}) ($p < 0.05$).

The reduction in SYE and EM under the \texttt{Understand} layer is more nuanced, especially when we focus on the combined variable-and-function renaming setting shown in Figure~\ref{rq2}. For ChatRepair, both renaming schemes consistently lower SYE and EM relative to the \texttt{Remember} baseline across all six model configurations. For example, $ChatRepair^{_{Remember}}_{_{GPT35}}$ drops from 17.05\% SYE and 9.68\% EM to 5.07\% SYE and 2.30\% EM under \texttt{Hash-based renaming}, and to 9.68\% SYE and 4.61\% EM under \texttt{Rephrasing-based renaming}. A similar pattern also appears for the stronger revised models: $ChatRepair^{_{Remember}}_{_{GPT54}}$ decreases from 17.97\% SYE and 17.51\% EM to 8.29\% SYE and 5.99\% EM under \texttt{Hash-based renaming}, and to 9.22\% SYE and 7.83\% EM under \texttt{Rephrasing-based renaming}. In contrast, CigaR remains more mixed. While the historical CigaR setups still show lower SYE and EM after renaming, some revised setups preserve similar or even slightly higher values than the \texttt{Remember} baseline. For instance, $CigaR^{_{Remember}}_{_{Qwen35Plus}}$ increases from 9.68\% SYE and 8.76\% EM to 11.06\% SYE and 10.60\% EM under \texttt{Hash-based renaming}, and remains at 11.06\% SYE and 10.14\% EM under \texttt{Rephrasing-based renaming}. These results indicate that lexical perturbations most consistently reduce the ability to obtain a plausible repair at all, whereas SYE and EM are more strongly shaped by the interaction between the underlying model and the APR agent. \respto{R1-11.1}

A direct comparison between the two renaming schemes further shows that neither one uniformly dominates the other across all setups and metrics. Under the combined renaming setting, rephrasing-based renaming usually preserves higher SYE and EM than \texttt{Hash-based renaming} for ChatRepair. For example, $ChatRepair^{_{Understand}}_{_{GPT35}}$ improves from 5.07\% to 9.68\% in SYE and from 2.30\% to 4.61\% in EM when moving from \texttt{Hash-based renaming} to \texttt{Rephrasing-based renaming}; similarly, $ChatRepair^{_{Understand}}_{_{StarCoder}}$ improves from 5.53\% to 9.22\% in SYE and from 4.61\% to 7.37\% in EM. CigaR exhibits a similar but not identical trend: rephrasing-based renaming yields higher SYE in five of the six setups and higher EM in most of them, such as $CigaR^{_{Understand}}_{_{GPT35}}$ rising from 5.07\% to 12.44\% in SYE and from 3.69\% to 5.53\% in EM, while $CigaR^{_{Understand}}_{_{Qwen35Plus}}$ remains nearly unchanged in SYE (11.06\% under both schemes) and is slightly lower in EM under rephrasing-based renaming (10.14\% versus 10.60\%). These results show that the \texttt{Understand} layer captures a broader sensitivity to current SUEs' identifier-level semantics, whether these semantics are removed entirely or transformed into alternative lexical cues.


These observations suggest that current LLM-powered APR solutions remain sensitive to surface-level lexical changes, even when the underlying bug logic is preserved. Across the evaluated setups, renaming-based perturbations consistently weaken repair performance, with the clearest and most stable effect appearing in PP. At the same time, some successful SYE and EM patches show that adaptation to the renamed context is still possible in certain cases. In this sense, the \texttt{Understand} layer is best interpreted as revealing limited and uneven lexical robustness, rather than a failure caused by changes in the underlying bug logic itself.\respto{R1-4.2}}


\mybox{\textbf{Summary}: \add{On \defectj, both \texttt{Hash-based renaming} and \texttt{Rephrasing-based renaming} substantially reduce repair performance at the \texttt{Understand} layer, with the clearest and most consistent drop appearing in PP across all twelve setups. The relative gap between the two renaming schemes is model-dependent and agent-dependent, but the overall trend is stable: current SUEs remain highly sensitive to identifier-level semantic cues. These results suggest that lexical robustness is still a major limitation, even when the underlying bug logic and execution behavior are preserved. In certain cases, the observed failures are consistent with the interpretation that current LLM-powered APR systems still depend heavily on surface-level textual cues, rather than consistently reasoning over the code in a way that is robust to identifier-level changes.}}

\subsection{Apply}
\label{sec:RQ3}
\add{Figure~\ref{rq2} reports the PP/SYE/EM results for the \texttt{Apply} layer on the synthetic bug-pattern dataset. Unlike the \texttt{Understand} layer, which perturbs lexical cues within the original benchmark instances, the \texttt{Apply} layer reconstructs new synthetic bugs, fixes, and tests in new local contexts while preserving similar buggy behavior patterns from the original benchmark.\respto{R1-3.1}}

\add{Overall, the \texttt{Apply} layer produces high repair rates across the twelve evaluated setups. PP ranges from 72.35\% to 94.47\%, while SYE and EM range from 18.43\% to 34.10\% and from 17.51\% to 33.64\%, respectively. The most notable result is the substantial gap between PP and the other two metrics, namely SYE and EM. Although the evaluated SUEs frequently generate patches that satisfy the synthetic tests, considerably fewer patches reproduce the intended repair syntactically or exactly. This gap can partly be explained by the construction of the \texttt{Apply} cases. The reconstructed instances are lightweight, single-file Java cases with limited dependencies and generated tests, making them comparatively easy to execute and satisfy. Consequently, the high PP rates primarily show that current SUEs can produce test-passing patches under controlled synthetic transfer. They do not necessarily indicate that the generated patches faithfully implement the intended repair behavior.}

\add{We further conducted a manual validation study on the quality of the generated PPs. We randomly sampled 50 bug IDs from the 217 single-function synthetic cases and reused the same sampled IDs across all 12 evaluated setups, resulting in 600 sampled cases in total. For each sampled bug and setup, we manually examined its first successfully generated PP. If no PP was generated for a given bug and setup, the case was marked as a failure. The twelve setups were divided equally between two authors, with each author reviewing 300 sampled cases. For each case, the assigned author examined the available metadata and repair outcome. If a PP was generated, the author further examined the intended behavior, the reference ground-truth code, and the generated patch to determine whether the PP genuinely repaired the underlying functionality rather than merely passing or bypassing the synthetic tests; otherwise, the case was marked as a failure.}

\begin{figure*}[!htbp]
    \centering
    \includegraphics[width=1\textwidth]{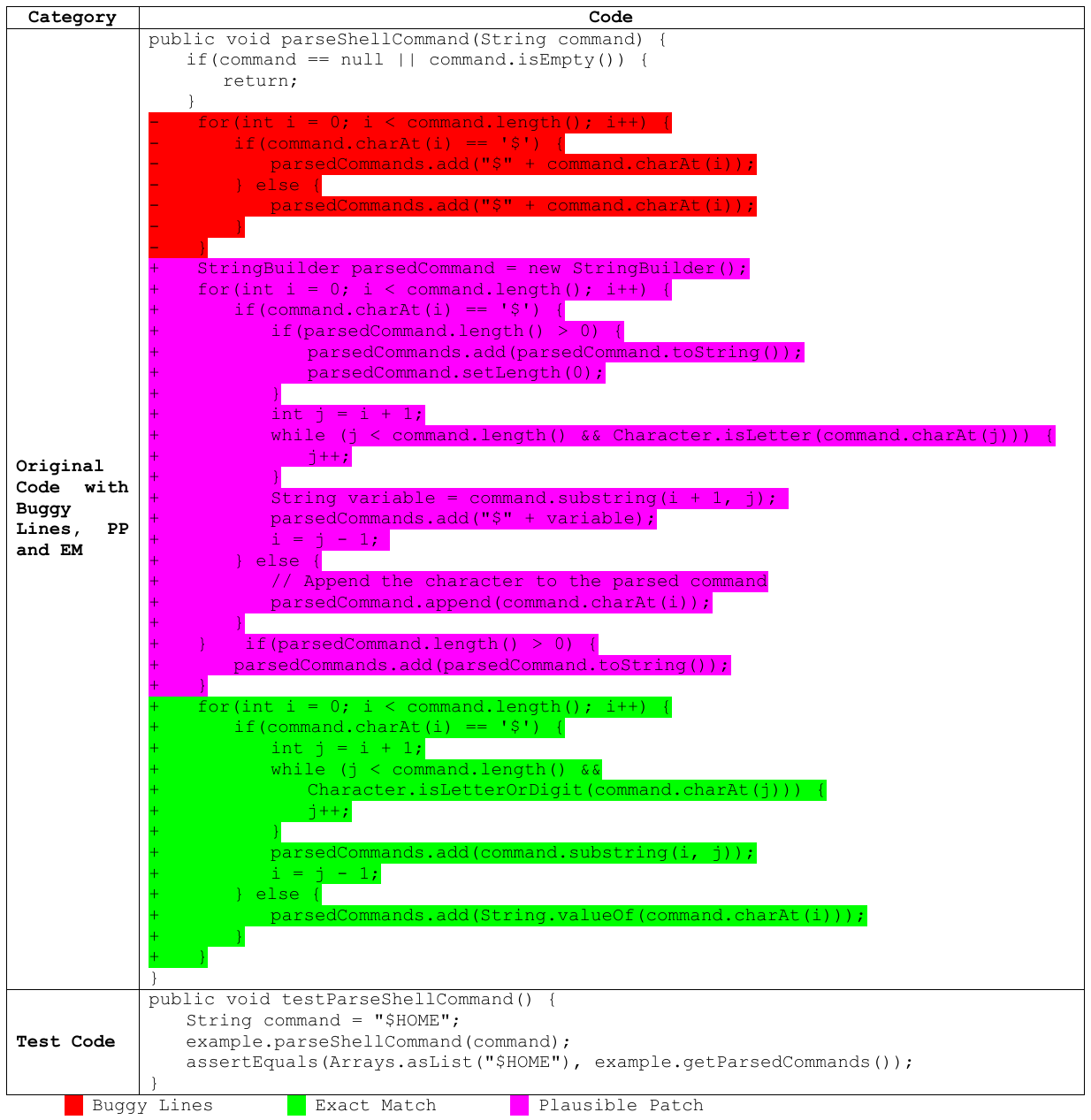}
    \caption{\normalsize \add{Case study: an example of Cli-2 with a PP generated by GPT35.}}
    \label{fig:case_s1}
\end{figure*}


\add{Figure~\ref{fig:case_s1} illustrates an example of a PP for \texttt{Cli-2} that passes the provided synthetic test but fails manual validation. The test covers only the input ``\$HOME'', for which the generated patch produces the expected output. However, the patch does not preserve the more general behavior of the reference fix. It accumulates consecutive non-variable characters into single strings, whereas the reference implementation records them as separate elements while grouping each environment-variable reference. Thus, for an input such as ``echo \$HOME/bin'', the generated and reference implementations produce different parsed representations. The patch also scans variable names using \texttt{Character.isLetter}, rather than \texttt{Character.isLetterOrDigit} as in the reference fix, causing variable references containing digits to be parsed differently. Because these discrepancies are not exposed by the narrow test for ``\$HOME'', automated evaluation classifies the patch as a PP even though the assigned manual reviewer classified it as an invalid repair.\respto{E0-3.1} \respto{R1-9.1}}


\begin{figure*}[t]
    \centering
    \includegraphics[width=1.00\textwidth]{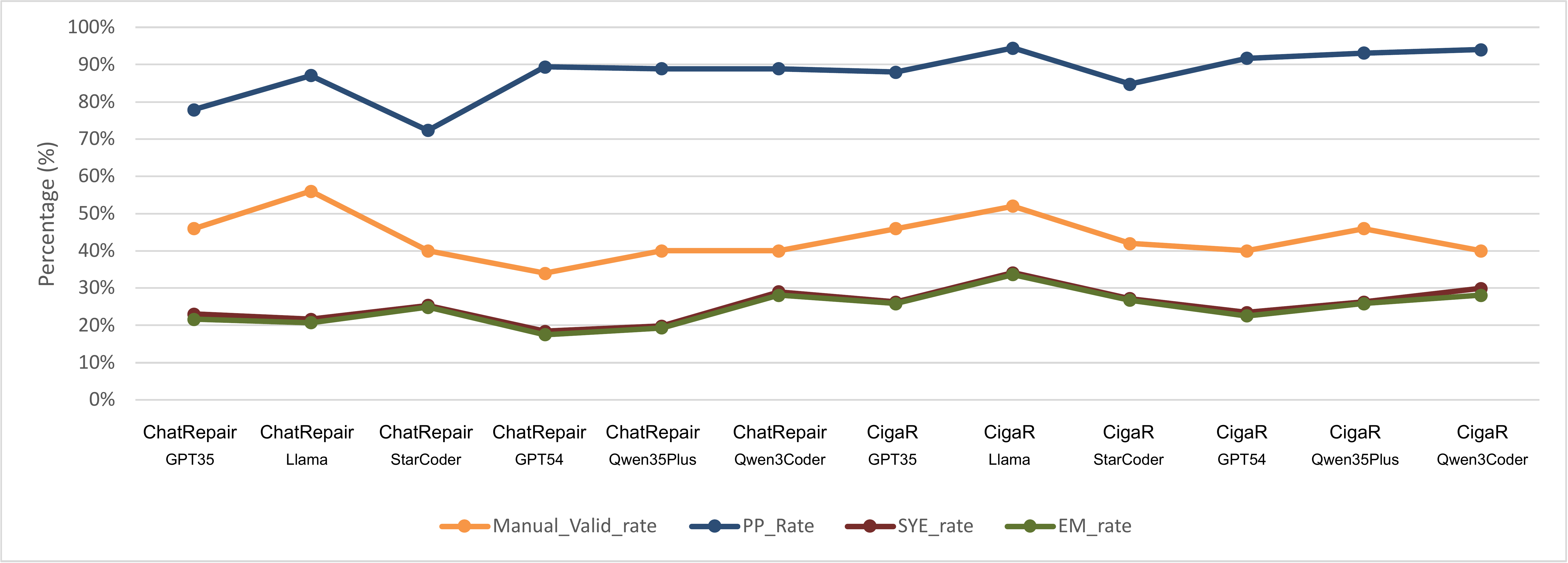}
    \caption{\normalsize \add{Sample-based manual-validity results on 50 shared bug IDs shown alongside full-dataset PP, SYE, and EM results from the 217-case \texttt{Apply} dataset for descriptive comparison.}}
    \label{fig:apply_manual_validity}
\end{figure*}

\add{Figure~\ref{fig:apply_manual_validity} presents the sample-based manual-validity results alongside the full-dataset PP, SYE, and EM results for descriptive context. Across the twelve setups, the end-to-end manually validated repair rate on the shared 50-ID sample ranges from 34.00\% to 56.00\%, with an average of 43.50\%. On the complete 217-case Apply dataset, PP ranges from 72.35\% to 94.47\%, with an average of 87.56\%.

These results use different denominators: manual validity is estimated from the shared 50-ID sample, whereas PP, SYE, and EM are calculated over all 217 Apply cases. We therefore do not interpret the numerical differences as matched percentage-point gaps, nor do we use the comparison to establish an exact validity rate for all generated PPs. Instead, the manual study serves as a sample-based assessment showing that test-passing patches may still fail to implement the intended repair behavior.\respto{R1-3.2}}

\add{The manual assessment confirms that passing the generated tests is not sufficient to establish semantic repair correctness in the \texttt{Apply} layer. Although the evaluated systems frequently produce PPs, a substantial proportion of these patches do not correctly implement the intended behavior. PP captures whether a patch satisfies the available tests, whereas manual validation provides a more direct assessment of whether the intended bug has actually been repaired. SYE and EM remain useful stricter measures, but they may reject semantically valid patches that differ from the reference fix.}

\add{We further compared ChatRepair and CigaR under matched LLM configurations. CigaR achieves higher PP than ChatRepair across all six matched model configurations and also shows higher overall SYE and EM ranges. ChatRepair achieves PP, SYE, and EM ranges of 72.35\%-89.40\%, 18.43\%-29.03\%, and 17.51\%-28.11\%, respectively, whereas CigaR achieves corresponding ranges of 84.79\%-94.47\%, 23.50\%-34.10\%, and 22.58\%-33.64\%. For example, under GPT54, ChatRepair reaches 89.40\% PP, 18.43\% SYE, and 17.51\% EM, compared with 91.71\%, 23.50\%, and 22.58\% for CigaR.}

\add{These differences show that APR-system design continues to influence repair performance even when the same underlying LLM is used. In the evaluated synthetic transfer settings, CigaR's multi-stage self-reflection-based workflow is associated with stronger repair results than ChatRepair. However, because we do not independently manipulate the individual architectural components, the observed differences should not be attributed solely to self-reflection.}

\add{Finally, the \texttt{Apply} results should be interpreted in relation to the other \tool layers. Compared with the memory-permissive \texttt{Remember} setting, the most pronounced increase at the \texttt{Apply} layer occurs in PP. This non-monotonic performance pattern does not imply that the cognitive capability represented by \texttt{Apply} is inherently easier than \texttt{Remember}. Instead, it reflects the lightweight, single-file construction and generated tests used in the current \texttt{Apply} layer instantiation. Compared with the \texttt{Analyze} layer, the \texttt{Apply} cases contain fewer dependencies, narrower behavioral scopes, and simpler local contexts. The results therefore characterize controlled transfer of repair knowledge into newly instantiated synthetic cases, rather than robust generalization to substantially different real-world project environments.}

\mybox{\textbf{Summary}: \add{On \defectj, the \texttt{Apply} layer shows that current SUEs can often transfer repair knowledge to newly synthetically generated bugs injected into simple project contexts, yielding substantially higher PP than the original benchmark. SYE and EM also increase in many configurations, although they remain far below PP and should be interpreted using paired configuration-level comparisons. On the synthetic bug-pattern dataset, PP ranges from 72.35\% to 94.47\%, SYE from 18.43\% to 34.10\%, and EM from 17.51\% to 33.64\%. However, strong performance on synthetic bugs should not be equated with robust generalization, as many PPs failed our manual validity assessment. Finally, these high repair rates do not necessarily reflect deep semantic understanding, as the reconstructed local contexts remain substantially simpler than the real-world project settings examined later in the \texttt{Analyze} layer. 
}}

\subsection{Analyze}
\label{sec:RQ4}
\begin{figure*}[t]
  \centering

    \begin{subfigure}[b]{0.33\textwidth}
        \includegraphics[width=\textwidth]{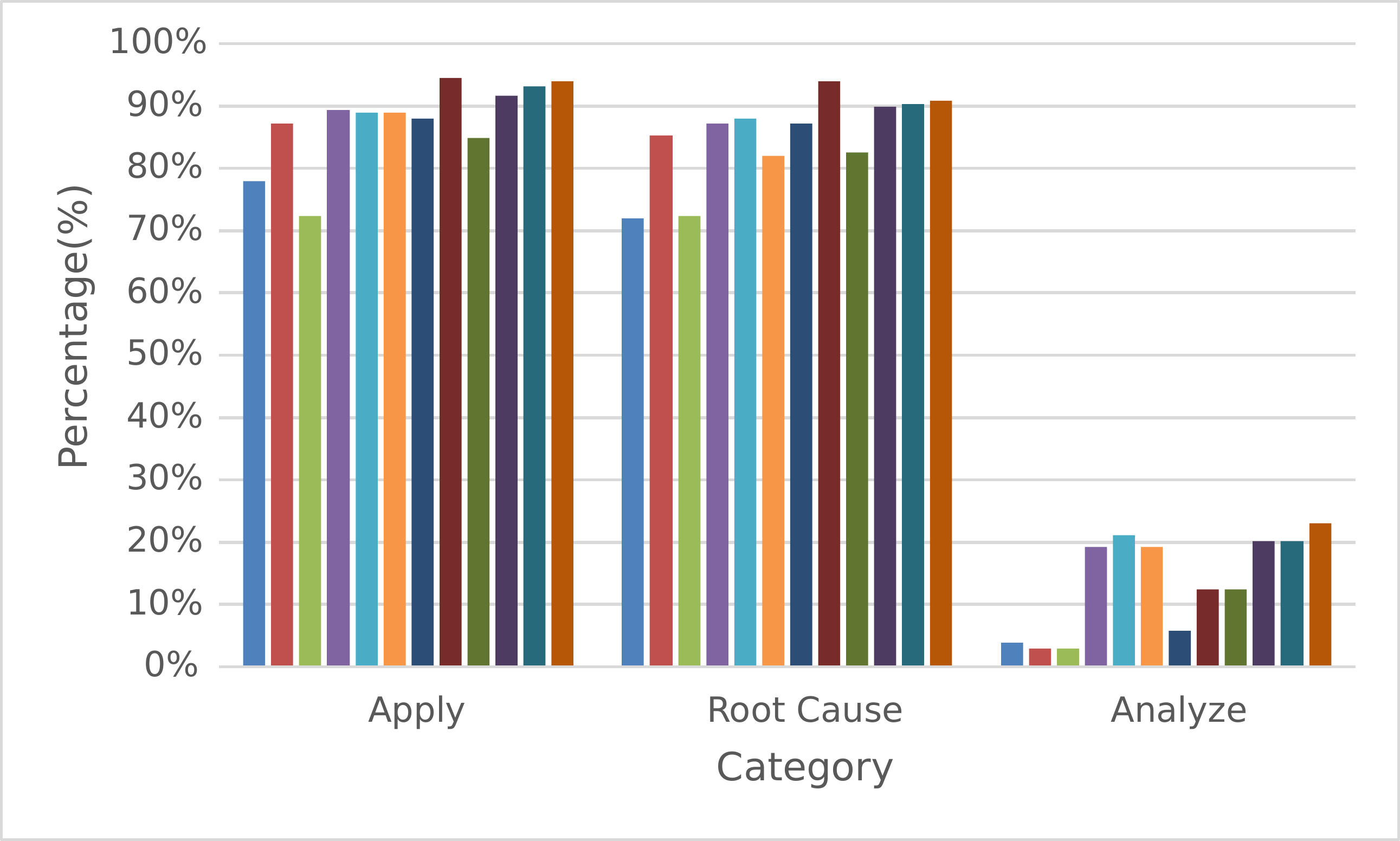}
        \caption{\footnotesize \% of Bugs with PP}
        \label{ppana}
    \end{subfigure}
    \hfill
    \begin{subfigure}[b]{0.33\textwidth}
        \includegraphics[width=\textwidth]{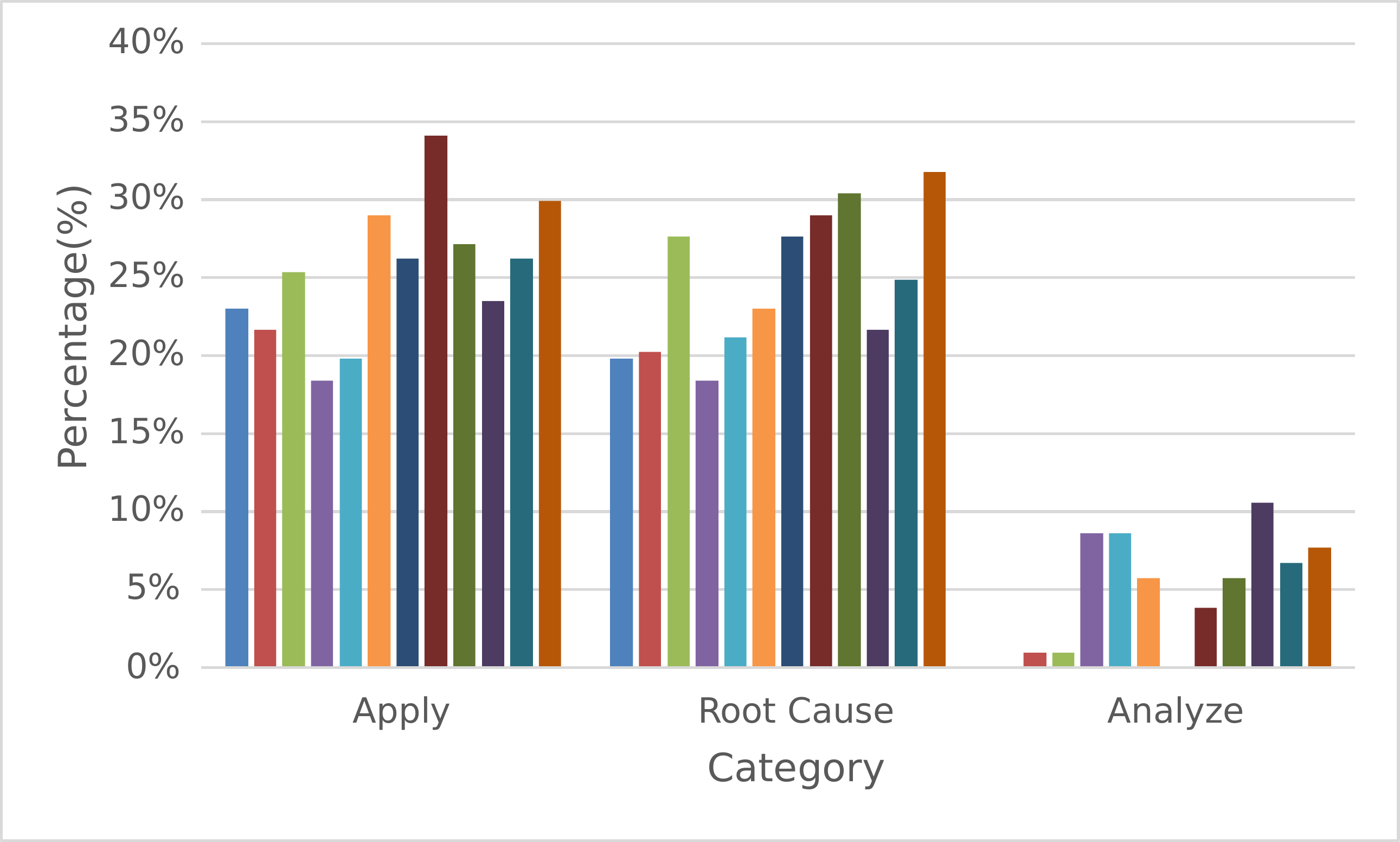}
        \caption{\footnotesize \% of Bugs with SYE}
        \label{syeana}
    \end{subfigure}
    \hfill
    \begin{subfigure}[b]{0.33\textwidth}
        \includegraphics[width=\textwidth]{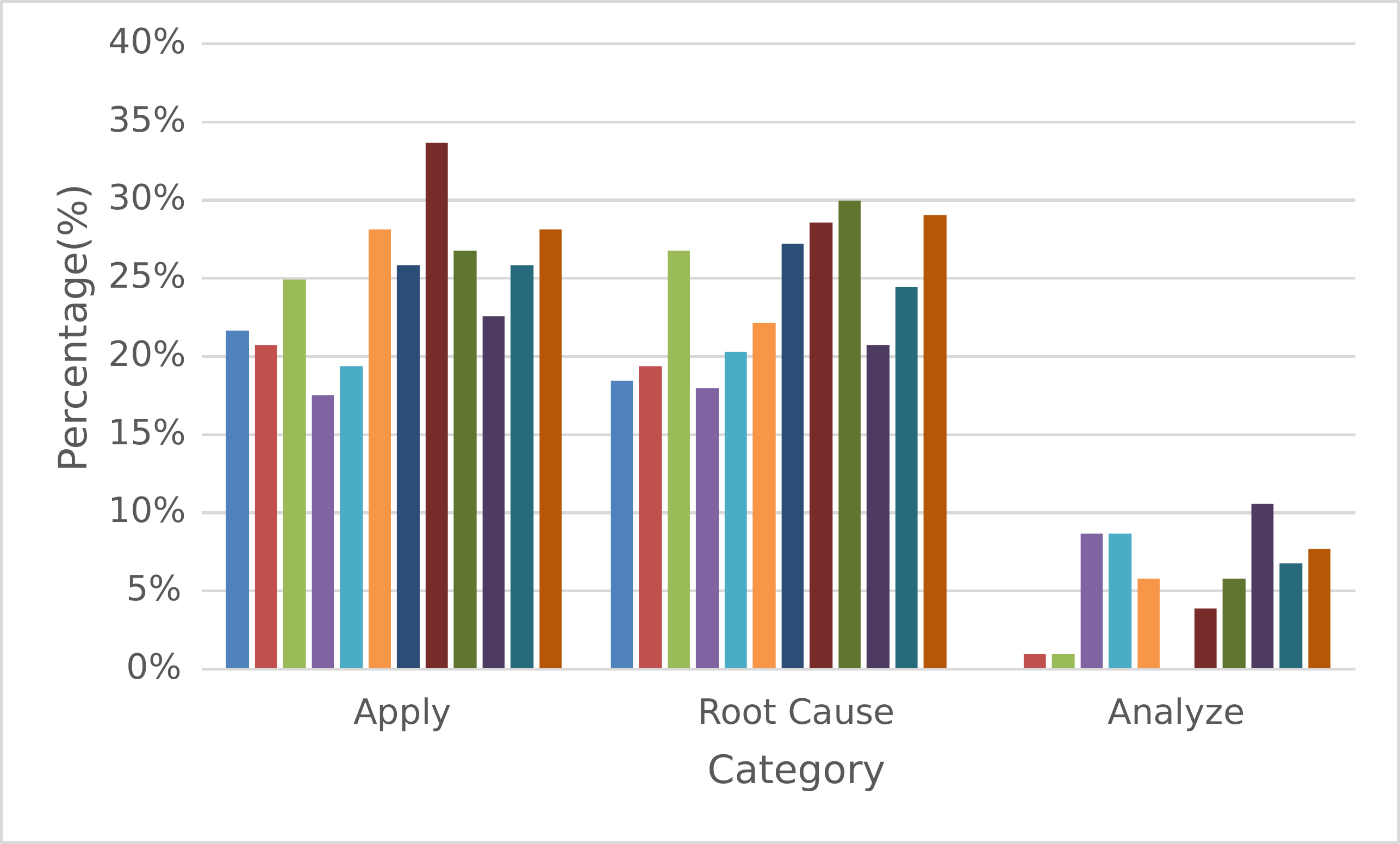}
        \caption{\footnotesize \% of Bugs with EM}
        \label{emana}
    \end{subfigure}
    \caption{\add{Aggregate PP, SYE, and EM evaluation results across three related \tool evaluation settings: the \texttt{Apply} layer ($N=217$), the synthetic bugs with the same root causes ($N=217$), and the real-world \texttt{Analyze} layer ($N=104$).\jack{fix the percentage number}\yinghang{Repeated figure for original Fig 10. Removed in Discussion section and kept here.}}}
    \label{compana}
\end{figure*}

\add{Figure~\ref{rq2} reports the PP/SYE/EM results for the \texttt{Analyze} layer on the finalized set of $104$ matched real-world cases. Unlike the lower layers, the \texttt{Analyze} layer no longer evaluates lexical perturbations or synthetic local reconstructions. Instead, it asks whether SUEs can identify and repair bugs that preserve the same underlying root cause as the benchmark instances while appearing in substantially different real-world project contexts.}

\add{Overall, the \texttt{Analyze} layer remains the most challenging setting in our evaluation. Across all twelve setups, PP ranges from 2.88\% to 23.08\%, SYE from 0.00\% to 10.58\%, and EM from 0.00\% to 10.58\%. These values are substantially lower than those in the \texttt{Apply} layer, showing that the transfer of bug-fixing knowledge becomes much harder once the bugs are embedded into realistic project environments with different dependencies, coding styles, and build settings.}

\add{Importantly, the real-world bugs in the \texttt{Analyze} layer are not intended to preserve the same surface buggy pattern, local implementation shape, or patch form as the original benchmark instances. Instead, the injection process is centered on preserving the same underlying root cause in a substantially different project environment. This design choice is deliberate: compared with the synthetic \texttt{Apply} layer, which still evaluates controlled transfer in relatively lightweight local contexts, the \texttt{Analyze} layer probes a deeper level of transfer, namely whether a SUE can recognize and repair the same failure mechanism when it is re-instantiated under different dependencies, coding styles, and project-specific structures.\respto{R1-5.1}}

\add{Compared with the lower layers, the \texttt{Analyze} layer introduces a fundamentally harder setting. The bugs now preserve the same underlying root causes as the benchmark instances, but they are embedded in substantially different real-world software environments. This shift introduces additional challenges, such as external dependencies, project-specific conventions, and more complex build and execution conditions. As a result, even when the same failure mechanism is preserved, transferring bug-fixing knowledge into a realistic project context remains difficult for current SUEs.}

\add{To better interpret the main real-world \texttt{Analyze} results, we further conducted an auxiliary comparison against two lighter-weight settings: the synthetically generated dataset from the \texttt{Apply} layer, which preserves the same bug patterns, and a newly generated synthetic dataset that shares the same root causes as the evaluated \defectj bugs. This comparison is intended as an auxiliary analysis for interpretation rather than as the primary definition of the \texttt{Analyze} layer itself. 

The three datasets have different sizes. The \texttt{Apply} and synthetic same-root-cause results are calculated over all 217 source bug IDs, whereas the real-world \texttt{Analyze} results are calculated over the 104-instance Analyze dataset for which validated root-cause-matched real-world cases could be identified and constructed. The smaller Analyze set reflects the practical availability of suitable real-world matches. Therefore, Figure~\ref{compana} should be interpreted as an aggregate cross-dataset comparison rather than as a paired bug-ID-level comparison.

\jiho{Is this comparison (Fig7) computed on the same 104 matched source bug IDs? The quoted Apply rates appear to be the full 217-case results, while Analyze uses 104 matched cases.}\yinghang{reversed the writing, see above and below.} Figure~\ref{compana} provides a comparison across three related \tool evaluation settings: the original \texttt{Apply} layer results, the synthetic same-root-cause evaluation results, and the real-world \texttt{Analyze} layer results. At the aggregate level, performance is substantially lower on the available real-world \texttt{Analyze} dataset than on the two full synthetic datasets. Under the original \texttt{Apply} layer, PP ranges from 72.35\% to 94.47\%, SYE from 18.43\% to 34.10\%, and EM from 17.51\% to 33.64\%. When we move to the synthetic dataset that preserves the same root cause but re-expresses the bug through a different bug pattern, the results remain relatively strong but become slightly weaker overall, with PP ranging from 71.89\% to 94.01\%, SYE from 18.43\% to 31.80\%, and EM from 17.97\% to 29.95\%. This intermediate result is important: even though the surface bug pattern has changed, the synthetic setting remains comparatively easy because the generated cases are still lightweight and local, typically involving only single-file contexts with limited dependencies. By contrast, once the same root-cause preservation is transferred into the real-world \texttt{Analyze}-layer setting, performance drops sharply: PP falls to 2.88\%-23.08\%, while both SYE and EM fall to 0.00\%-10.58\%. This descriptive contrast is consistent with the interpretation that realistic project contexts introduce substantial additional difficulty beyond lightweight synthetic reconstruction. However, because the \texttt{Analyze} dataset contains only the 104 instances for which suitable real-world matches could be obtained, differences in dataset size and case availability may also contribute to the observed aggregate gap. We therefore do not interpret Figure 7 as a paired or causally controlled comparison.\respto{R1-2.4} \respto{R1-5.2}}

\add{For PP, the strongest configuration is $CigaR^{_{Analyze}}_{_{Qwen3Coder}}$, which reaches 23.08\%, whereas the weakest are $ChatRepair^{_{Analyze}}_{_{Llama3}}$ and $ChatRepair^{_{Analyze}}_{_{StarCoder}}$, both at 2.88\%. For SYE and EM, $CigaR^{_{Analyze}}_{_{GPT5}}$ reaches the highest value at 10.58\%, while both $ChatRepair^{_{Analyze}}_{_{GPT}}$ and $CigaR^{_{Analyze}}_{_{GPT}}$ yield 0.00\%. These results suggest that real-world repair performance remains strongly dependent on both the model backbone and the APR agent, and that stronger coding-oriented backbones can improve performance substantially without removing the large gap between synthetic and real-world transfer.}


\mybox{\textbf{Summary}: \add{On \defectj, the \texttt{Analyze} layer remains the most challenging evaluation setting in \tool, because it requires SUEs to transfer bug-fixing knowledge while preserving the same underlying root causes in substantially different real-world project environments. On the finalized set of $104$ matched real-world cases, across all twelve setups, PP ranges from 2.88\% to 23.08\%, SYE from 0.00\% to 10.58\%, and EM from 0.00\% to 10.58\%. Considered together with the aggregate descriptive results from the full Apply dataset and the full synthetic same-root-cause dataset, the available 104-case Analyze results are consistent with the interpretation that realistic project contexts introduce additional difficulty beyond lightweight synthetic reconstruction. However, because these datasets differ in size and case composition, this comparison is not paired or causally controlled, and the observed aggregate gap may also be influenced by the availability and selection of root-cause-matched real-world cases.}}

\section{\SWE Study}
\label{sec:swebench}

\add{Although we demonstrated our research framework on \defectj using two representative APR systems (ChatRepair and CigaR) in the previous section, \tool is not restricted to a single static benchmark or programming language. To examine whether the framework can also be instantiated beyond Java and \defectj, we conducted a second case study using the \SWE snapshot accessed in April 2026, which contained 300 bug instances. We applied an automated filtering script to each instance's gold patch. The script identified the files and functions modified by the gold patch and retained only cases whose required change was localized to a single function in a single file. This procedure excluded 110 multi-function and/or multi-file cases and yielded 190 single-function instances.

Unless otherwise stated below, we constructed the \texttt{Remember}, \texttt{Understand}, \texttt{Apply}, and \texttt{Analyze} datasets using the same \tool pipeline described for the \defectj study in Section~\ref{sec:framework}. We evaluated three current LLMs: GPT54, Qwen35Plus, and Qwen3Coder, a subset of the six models used in the \defectj study. This choice accounts for the deprecation or unavailability of several older models by the time of this evaluation, while aligning with the study's primary objective of supplementary cross-benchmark validation. Claude-4.6-Sonnet was used to generate the corresponding synthetic and injected variants as Claude-3.5-Sonnet was no longer available. For the construction of the \texttt{Analyze} layer bugs, we used OpenAI Codex CLI~\cite{openaicodexcli} as the LLM-assisted code-search tool because it supported the repository-aware and web-assisted retrieval needed to identify matched real-world contexts. \jiho{minor: maybe a brief setup details for \SWEA?} We used \SWEA as the APR system because it is designed for repository-level \SWE tasks, whereas ChatRepair and CigaR are tightly coupled to their original benchmark-oriented workflows. We ran \SWEA using its default configuration without additional task-specific parameter tuning. For each case, \SWEA operated on the corresponding \SWE repository and was evaluated using the original or BloomAPR-modified tests associated with that case. A generated patch was counted as a PP if it passed all tests used for the corresponding evaluation instance. SYE and EM were calculated from the resulting patch using the same comparison procedures described in Section~\ref{sec:framework}. 

Since this case study differs from the \defectj study in programming language, APR agent, available model set, generator version, code-search tool, and repository context, it is not intended as a controlled project-level replication. Instead, it assesses whether \tool and its broad layer-wise evaluation pattern can be instantiated on a Python benchmark under a contemporary repository-level APR setting. \respto{E0-2.5} \respto{R1-6.2} \respto{R2-3.1} \yinghang{Rewrite the whole above to address multiple concerns for SWE-bench experiment.}}

\begin{figure*}[t]
  \centering

    \begin{subfigure}[b]{0.33\textwidth}
        \includegraphics[width=\textwidth]{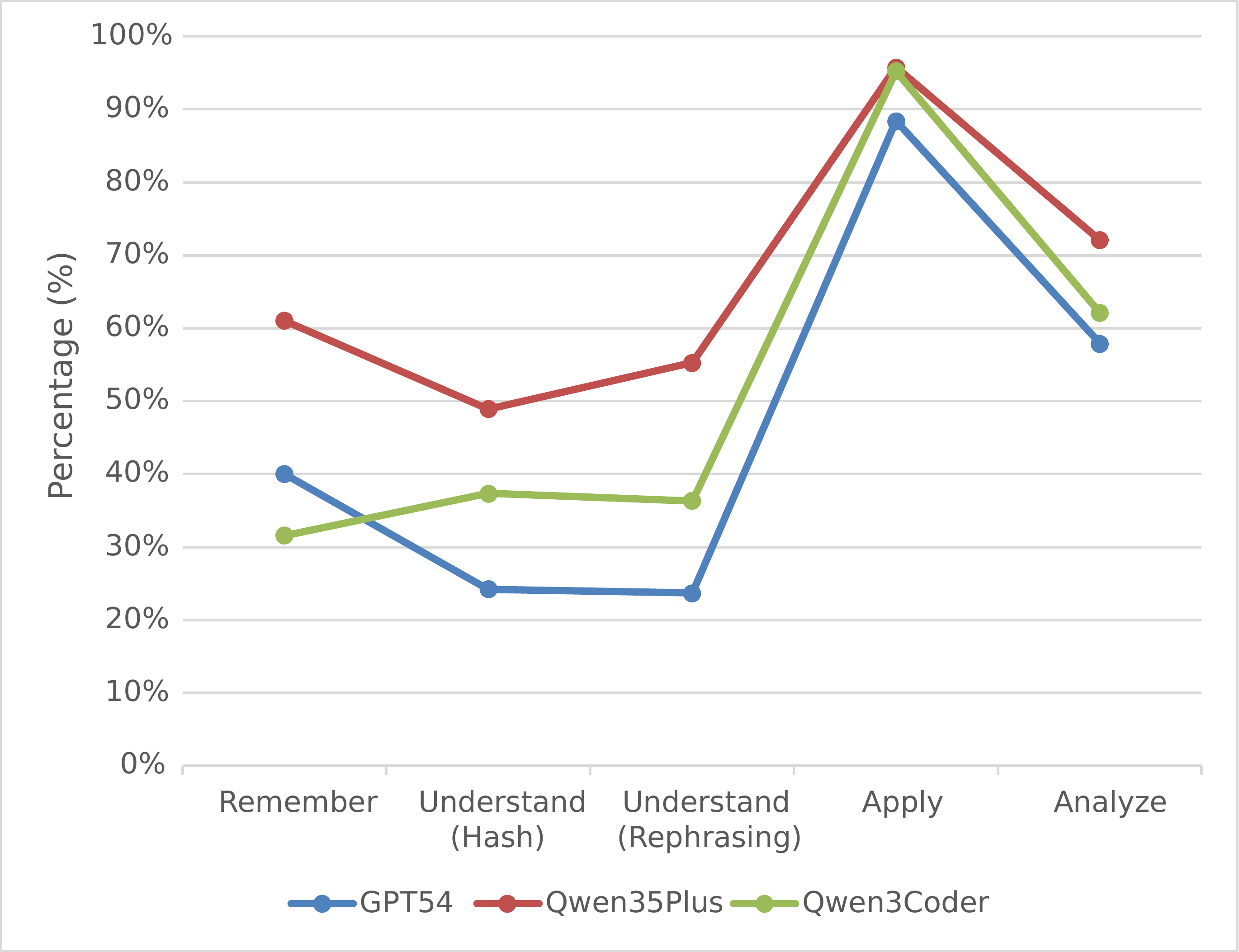}
        \caption{\footnotesize \% of Bugs with PP}
        \label{rq2swe_pp}
    \end{subfigure}
    \hfill
    \begin{subfigure}[b]{0.33\textwidth}
        \includegraphics[width=\textwidth]{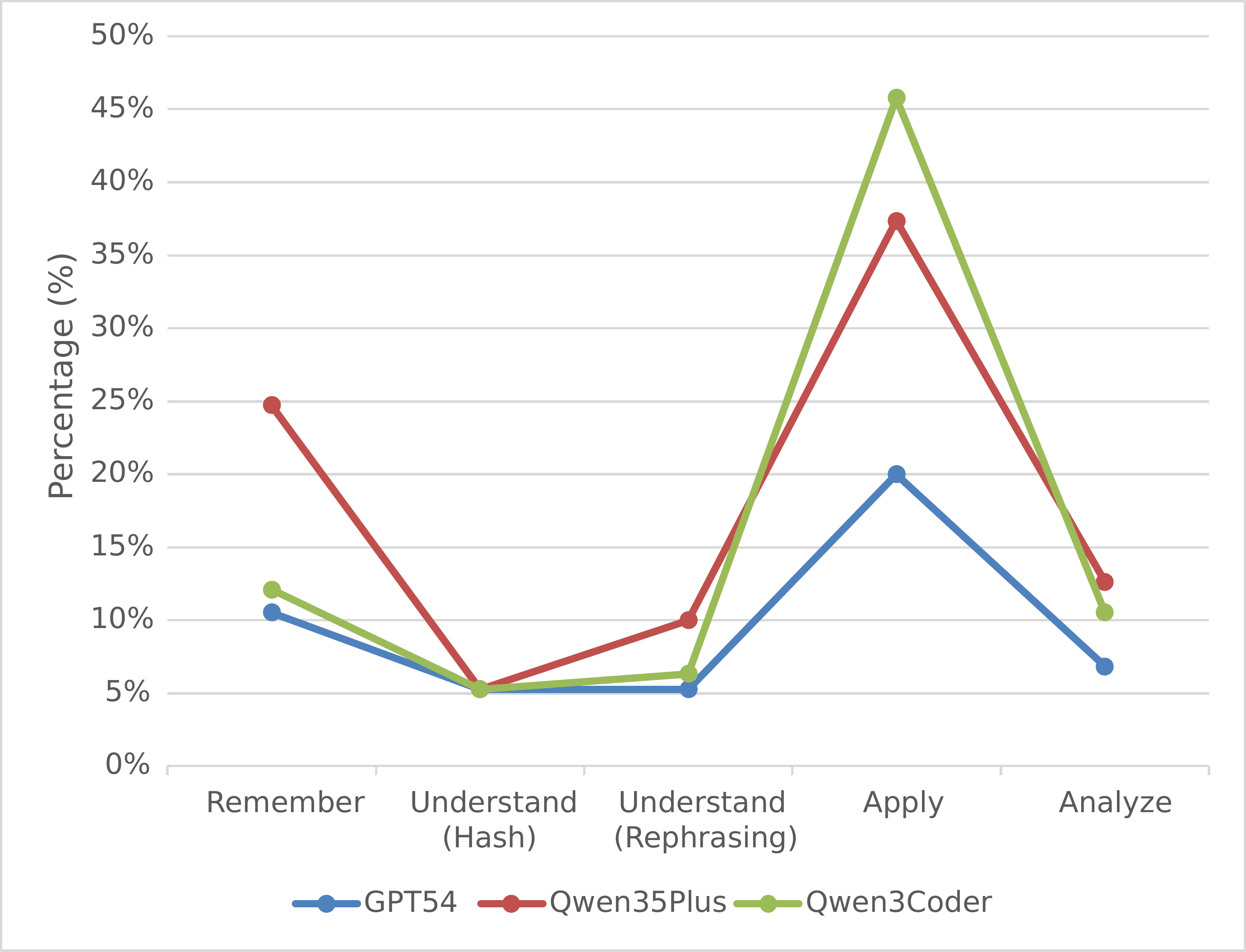}
        \caption{\footnotesize \% of Bugs with SYE}
        \label{rq2swe_sye}
    \end{subfigure}
    \hfill
    \begin{subfigure}[b]{0.33\textwidth}
        \includegraphics[width=\textwidth]{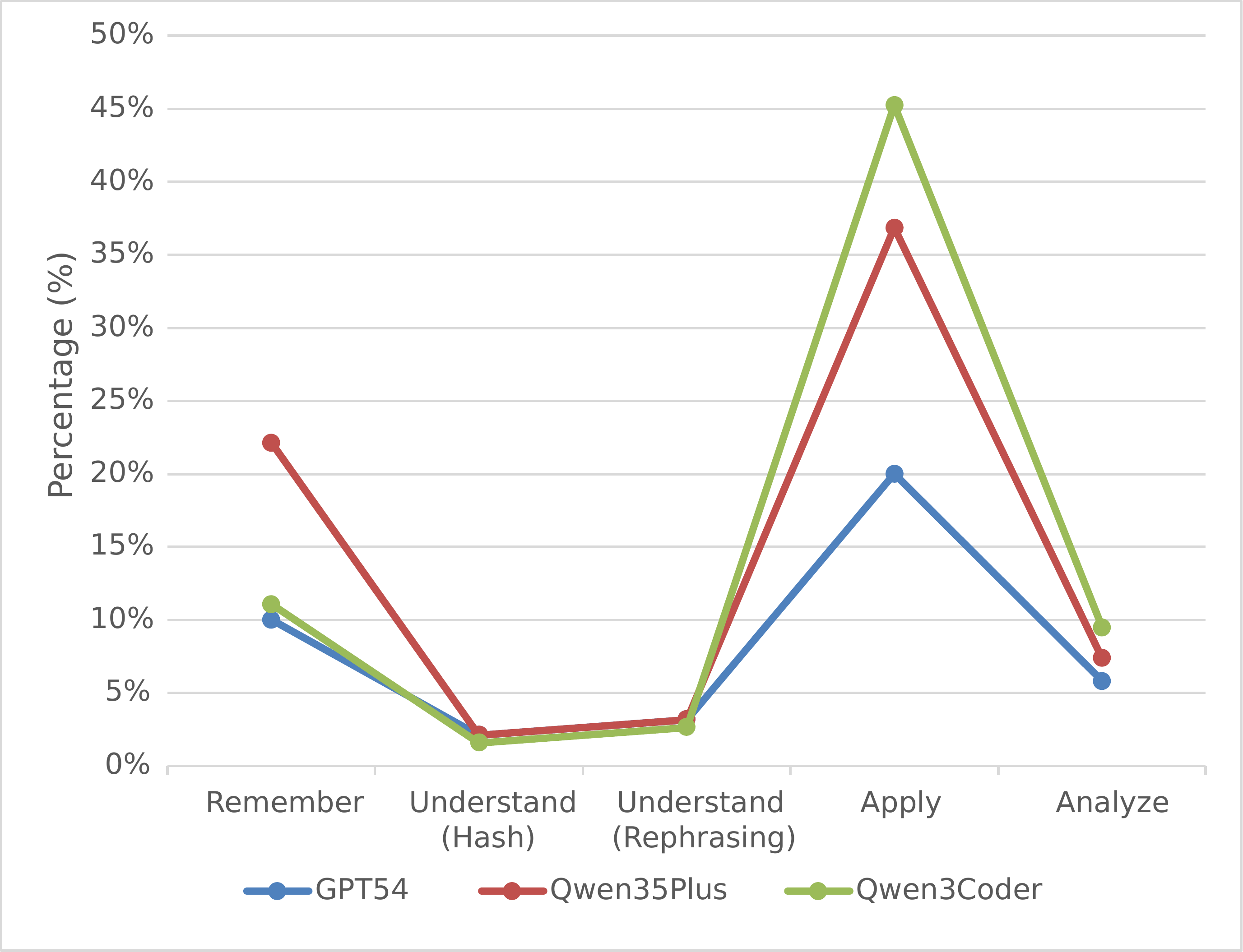}
        \caption{\footnotesize \% of Bugs with EM}
        \label{rq2swe_em}
    \end{subfigure}
    \caption{\add{PP, SYE, and EM results across \tool layers on \SWE.}}
    \label{rq2swe}
\end{figure*}

\add{Figure~\ref{rq2swe} presents the corresponding PP/SYE/EM results across the four layers for the 190 selected single-function cases from \SWE. At the \texttt{Analyze} layer, the selected cases are mapped to 190 real-world cases in a one-to-one manner, distributed across 148 real-world projects from GitHub.}

\add{At the \texttt{Remember} layer, as shown in Figure~\ref{rq2swe}, the \SWE results yield PP rates ranging from 31.58\% to 61.05\%, SYE rates from 10.53\% to 24.74\%, and EM rates from 10.00\% to 22.11\%. Among the three models, Qwen3.5-Plus achieves the strongest performance across all three metrics. Although the absolute values differ from those on \defectj, the \SWE results likewise show that plausible repair is considerably easier than exact reproduction, supporting our use of the \texttt{Remember} layer as a baseline under possible prior exposure rather than as evidence of guaranteed exact recall.}

\add{At the \texttt{Understand} layer, the overall trend remains consistent with the \defectj study: lexical perturbations substantially reduce repair performance. Under \texttt{Hash-based renaming}, PP ranges from 24.21\% to 48.95\%, whereas under \texttt{Rephrasing-based renaming} it ranges from 23.68\% to 55.26\%. The SYE rate is 5.26\% for each of the three LLMs under \texttt{Hash-based renaming}. Under \texttt{Rephrasing-based renaming}, SYE ranges from 5.26\% to 10.00\%. The EM rate ranges from 1.58\% to 2.11\% and from 2.63\% to 3.16\%, respectively. One notable difference appears for Qwen3-Coder, whose PP performance remains nearly unchanged between \texttt{Hash-based} and \texttt{Rephrasing-based renaming} (37.37\% versus 36.32\%).} 

\add{The \texttt{Apply} layer on \SWE also follows the same broad trend as in the \defectj study: synthetically reconstructed cases are considerably easier to resolve than cases at the \texttt{Remember} and \texttt{Analyze} layers. PP ranges from 88.42\% to 95.79\%, while SYE ranges from 20.00\% to 45.79\% and EM ranges from 20.00\% to 45.26\%. These results are again consistent with the \defectj study, suggesting that the synthetic bugs from the \texttt{Apply} layer capture controlled semantic transfer in relatively lightweight local contexts, which are easier to resolve than bugs in realistic project environments.}

\add{At the \texttt{Analyze} layer, the selected \SWE cases yield PP rates from 57.89\% to 72.11\%, SYE rates from 6.84\% to 12.63\%, and EM rates from 5.79\% to 9.47\%. These values remain substantially lower than the results from the \texttt{Apply} layer. Interestingly, however, the PP range is generally higher than the corresponding \defectj study at the \texttt{Analyze} layer. A plausible explanation is that the \SWE evaluation is conducted using SWE-agent, which has access to the full repository context. This provides much richer project-level context for APR diagnosis and repair. In contrast, the ChatRepair and CigaR experiments in our \defectj study operate on extracted \defectj-oriented metadata and do not have equivalent access to the full codebase. As a result, the \SWE real-world results should be interpreted as reflecting both the benchmark difference and the stronger contextual support available to the agent.}

\mybox{\textbf{Summary}: \add{The \SWE results remain broadly consistent with the \defectj study in terms of cross-layer trends (e.g., the performance drop from \texttt{Remember} to \texttt{Understand}, and from \texttt{Apply} to \texttt{Analyze}). 
Because this case study uses a different programming language, APR agent, model subset, generator version, search tool, and repository context, it should be interpreted as cross-benchmark validation rather than as a controlled project-level replication. \yinghang{Add a consistent note here.}
At the same time, they highlight two notable differences. First, the gap between \texttt{Hash-based renaming} and \texttt{Rephrasing-based renaming} is smaller for Qwen-3-Coder on the selected \SWE cases, suggesting greater robustness to identifier-level perturbations in this Python-based  \SWE setting. Second, real-world PP results are higher in the SWE-agent settings, potentially because the SWE-agent has access to richer repository-level context than the SUEs used in the \defectj study. Together, these findings provide preliminary evidence that the four-layer framework can be instantiated beyond a single benchmark and programming language via \tool. However, quantitative performance remains sensitive to the choice of benchmark, programming language, and the scope of context provided to the APR system.\respto{R2-2.2}}}
\section{Discussion and Results Implications}
\label{sec:discussion}

\subsection{Assessment Summary}
\label{sec:assessmentsum}


\add{Our case studies show that current LLM-powered APR solutions can benefit substantially from benchmark-familiar bug-fixing patterns. However, this advantage does not translate into robust performance when evaluation extends beyond the static benchmark settings. In several instances, the studied APR systems continue to generate repairs that mirror benchmark-familiar solutions rather than adapting to modified local context. For example, both $ChatRepair_{GPT35}$ and $CigaR_{GPT35}$ repeatedly produced identical fixes for the \texttt{Compress-7} bug from \defectj, even when function names were renamed at the \texttt{Understand} layer. Although both SUEs employ iterative refinement using feedback mechanisms like compilation and test execution, these workflows proved ineffective as the agents continuously re-introduced original function names. Furthermore, across both \defectj and \SWE studies, performance decreases from the synthetic Apply setting to the real-world Analyze setting, particularly for faithful-repair metrics. This suggests limited robustness in the evaluated pipelines when moving from controlled local contexts to broader project environments. These results highlight two critical limitations of current LLM-powered APR systems: (1) sensitivity to surface-level text and benchmark-familiar code fragments, which can lead systems to reproduce familiar repairs that fail after the local context is altered\respto{R1-12.1}, and (2) limited context-scaling capability when repair requires broader repository information rather than localized bug snippets~\cite{li2024longiclbench}. \respto{R1-12.2} Novel techniques are therefore required to improve context adaptation and long-context reasoning across both model- and agent-level architectures.}


\subsection{Patch Assessment in the Context of LLM-powered APR Solutions}
\label{sec:patchassessment}
Many existing LLM-powered APR systems, including the evaluated ChatRepair, CigaR, and \SWEA systems, primarily validate generated patches against the available tests. Passing those tests alone does not establish semantic correctness. 
Moreover, our manual analysis of the sampled \texttt{Apply} layer cases in the \defectj study reveals that many sampled PPs generated by these APR solutions failed manual validation and were invalid. In practice, validating whether a patch is truly correct remains challenging and may require stronger or additional tests, semantic analysis, or human review, and our findings suggest that this process cannot yet be fully automated by current LLMs alone. Unfortunately, relatively little prior patch-assessment and selection work  (e.g.,~\cite{zhou2023patchzero,le2023invalidator,le2019reliability}) has specifically examined patches generated by LLM-powered APR systems, whose characteristics can differ substantially from those of rule-based or deep-learning-based APR systems. Further research is needed in this area in order to ensure LLM-powered APR systems are safe for practical use.

\subsection{Static Benchmark Adaptation Strategies}
\label{sec:discussion_layer_generalization}

\add{The layered design of \tool should be understood as an evaluation methodology rather than as a fixed list of mutation operators. Each layer is defined by the type of capability it is intended to assess, while the concrete benchmark adaptations used in this paper represent one research-prototype instantiation of that layered design. This distinction is important for interpreting both our results and the scope of the framework. Our \defectj and \SWE studies demonstrate that the layered evaluation principle can be instantiated concretely and can already reveal important weaknesses of current LLM-powered APR systems. At the same time, the value of the framework does not depend on any single transformation recipe. The key idea is that each layer captures a progressively deeper form of repair capability, and therefore admits multiple valid realizations as summarized in Table~\ref{tab:layer_generalization}.
In the current paper, \texttt{Remember}, \texttt{Understand}, \texttt{Apply}, and \texttt{Analyze} are instantiated using the concrete transformations described in Section~\ref{sec:framework}, but these are not intended to exhaust the design space. Rather, they serve as representative realizations that demonstrate feasibility. This perspective also helps clarify the roles of the later Bloom layers, \texttt{Evaluate} and \texttt{Create}, which are not yet fully operationalized in our current prototype but provide natural directions for future benchmark evolution.\respto{E0-2.1}\respto{R1-1.4}\respto{R2-2.2}} 

\begin{table*}[t]
\centering
\small
\begin{tabular}{p{1.4cm} p{3.3cm} p{3.3cm} p{5.4cm}}
\toprule
\textbf{Layer} & \textbf{Evaluation Goal} & \textbf{Current Instantiation in \tool} & \textbf{Other Possible Realizations} \\
\midrule
\texttt{Remember}
& Reusing the existing static APR benchmark
& Original benchmark instances
& \multicolumn{1}{c}{--} \\
\hline
\texttt{Understand}
& Modifying the static benchmark using semantically equivalent code transformations
& Hash-based and rephrasing-based renaming of identifiers
& \begin{tabular}[t]{@{}l@{}}
1. Structural changes such as method reordering \\
2. Surface-level code changes such as adding or \\
\ \ removing redundant code or comments \\
3. Injecting code smells or refactoring \\
4. Control-flow changes or equivalent \\ 
\ \ API/library substitutions
\end{tabular} \\
\hline
\texttt{Apply}
& Synthesizing new bugs that mirror the bug-fixing patterns of the original static benchmark
& Synthetic bug-pattern variants in lightweight local contexts
& \begin{tabular}[t]{@{}l@{}}
1. Similar bugs across different programming  \\
\ \  languages or newer language versions \\
2. Multiple synthetic variants \\
\ \  for the same bug pattern
\end{tabular} \\
\hline
\texttt{Analyze}
& Injecting bugs that preserve the root causes of the original static benchmark into real-world projects
& One-to-one matched real-world cases with the same underlying root cause
& \begin{tabular}[t]{@{}l@{}}
1. Different bug patterns but the \\
\ \ same root cause within the same project \\
2. The same underlying root cause \\
\ \ re-instantiated through different surface \\
\ \ bug patterns across multiple project contexts.
\end{tabular} \\
\hline
\texttt{Evaluate}
& Assessing the correctness of the generated patches
& Not fully instantiated in the current prototype
& \begin{tabular}[t]{@{}l@{}}
1. Stronger PP evaluation with better testing or \\
\ \ richer criteria \\
2. Repair validation without pre-existing tests, \\
\ \ requiring test generation for fix verification
\end{tabular} \\
\hline
\texttt{Create}
& Generating new bug repair tasks with previously unseen characteristics
& Not fully instantiated in the current prototype
& \begin{tabular}[t]{@{}l@{}}
1. Bugs with unseen bug-fixing patterns or \\
\ \ workflows \\
2. Newly curated or compound bugs within the \\
\ \ same project context
\end{tabular} \\
\bottomrule
\end{tabular}
\caption{\add{The realizations for \tool across different layers.}}
\label{tab:layer_generalization}
\end{table*}

\add{\subsubsection{\texttt{Understand}} \hfill\\
At the \texttt{Understand} layer, the central question is whether an APR system can still construct the intended code meaning when the source-level presentation is perturbed without changing the underlying bug logic or execution behavior. In our prototype, we operationalized this layer primarily through identifier renaming, including both \texttt{Hash-based renaming} and \texttt{Rephrasing-based renaming}. However, the methodological scope of this layer is broader than renaming alone. It can naturally include the following additional realizations: \respto{E0-2.3}

\begin{itemize}[leftmargin=*]
    \item \textbf{Structural changes}: reordering methods or reorganizing source-level structure while preserving the intended repair target.
    \item \textbf{Surface-level code changes}: adding or removing redundant code, comments, or other semantically irrelevant presentation cues that do not alter the execution behavior.
    \item \textbf{Code-smell or refactoring injections}: introducing local refactoring patterns or code smells that change readability or organization without changing the underlying defect logic.
    \item \textbf{Control-flow or API-level rewriting}: replacing one equivalent local implementation with another, such as changing a \texttt{for} loop to a \texttt{while} loop or substituting equivalent APIs or libraries, provided that the bug and its intended fix remain the same.
\end{itemize}

The methodological purpose of this layer is therefore not to restrict evaluation to a single operator, but rather to assess whether an APR system can accurately interpret code when superficial presentation cues are altered. To this end, identifier renaming, method reordering, redundant-code insertion or removal, comment-level modifications represent valid realizations of this core evaluation objective.}

\add{\emph{Pilot Study}: 
To illustrate this broader interpretation, we conducted an exploratory pilot study using comment-based perturbations. We first used an automated script to identify eligible \defectj bugs satisfying two conditions: (1) the original, unperturbed bug produced a PP under all 12 SUE/model configurations, and (2) its source code contained both textual comments and code comments. We then randomly selected three bugs from this filtered set. Here, a textual comment refers to a natural-language comment containing no executable code fragment, whereas a code comment contains an actual code fragment.\yinghang{Reworded to be clear} For each bug, we constructed four semantics-preserving variants by (1) adding textual comments, (2) removing textual comments, (3) adding code comments, and (4) removing code comments, while leaving the underlying bug logic, repair target, and runtime behavior unchanged. Across the resulting 12 perturbed scenarios and 12 SUEs (ChatRepair and CigaR under six LLM backbones), this yielded 144 evaluation runs in total. All results reported below are PP-level outcomes: a run is counted as repairable if it produces a PP; otherwise, it is counted as unsuccessful. Among them, 74 runs still produced PPs, whereas 70 did not, showing that even non-functional comment-level perturbations alone can materially affect APR behavior.

At the PP level, the added-textual-comment setting is the least disruptive, with 22/36 runs (61.11\%) remaining repairable, followed by removing code comments, with 18/36 runs (50.00\%) remaining repairable. Removing textual comments and adding code comments are slightly more disruptive, each retaining repairability in 17/36 runs (47.22\%). Grouped more coarsely, textual-comment perturbations preserve repairability in 39/72 runs (54.17\%), whereas code-comment perturbations preserve repairability in 35/72 runs (48.61\%). Similarly, addition-based perturbations preserve repairability in 39/72 runs (54.17\%), whereas removal-based perturbations preserve repairability in 35/72 runs (48.61\%). These differences are not large enough for us to claim a universal ranking across all comment operators, but they show that semantics-preserving comment manipulations can change PP outcomes in a substantial proportion of the evaluated runs.

The effect on PP outcomes is also strongly bug-dependent. \texttt{Gson-5} remains comparatively robust, with 40/48 perturbed runs (83.33\%) producing PPs. \texttt{JacksonDatabind-17} shows moderate sensitivity, with 29/48 perturbed runs (60.42\%) producing PPs. In contrast, \texttt{JacksonDatabind-85} is highly sensitive, with only 5/48 perturbed runs (10.42\%) producing PPs overall, including no PP under the added-code-comment condition (0/12, 0.00\%). Taken together, these exploratory PP-level results support the broader interpretation of the \texttt{Understand} layer: current APR systems are sensitive not only to identifier renaming, but also to semantics-preserving changes in how buggy code is presented.
\respto{R2-2.1}}

\begin{figure*}[t]
  \centering
  \includegraphics[width=0.7\textwidth]{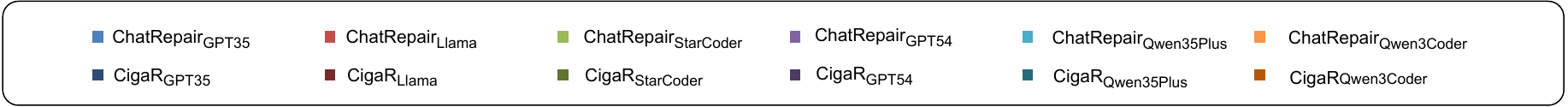}

    \begin{subfigure}[b]{0.33\textwidth}
        \includegraphics[width=\textwidth]{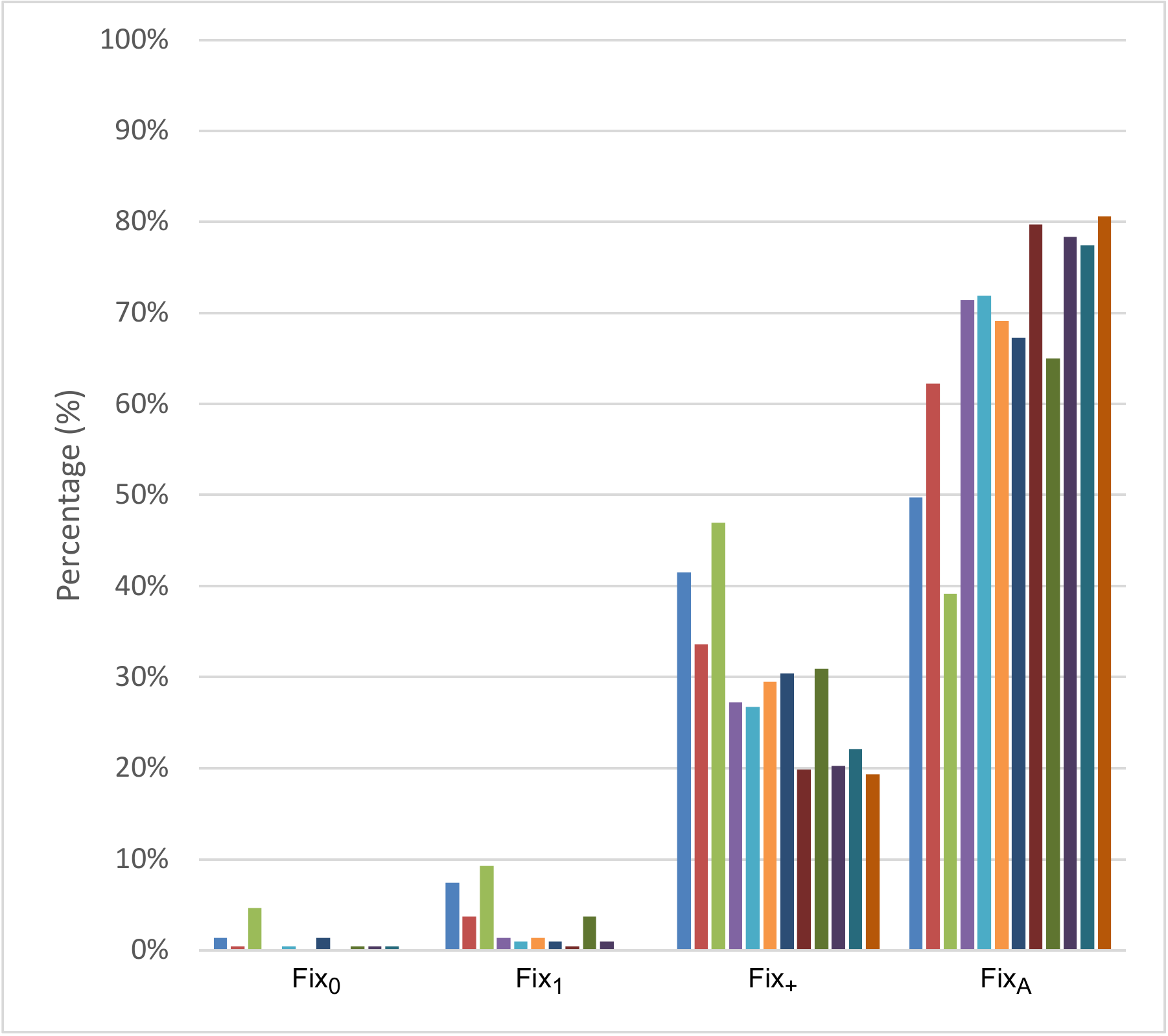}
        \caption{\footnotesize \% of Bugs with PP}
        \label{fig:pp_apply_consistency}
    \end{subfigure}
    \hfill
    \begin{subfigure}[b]{0.33\textwidth}
        \includegraphics[width=\textwidth]{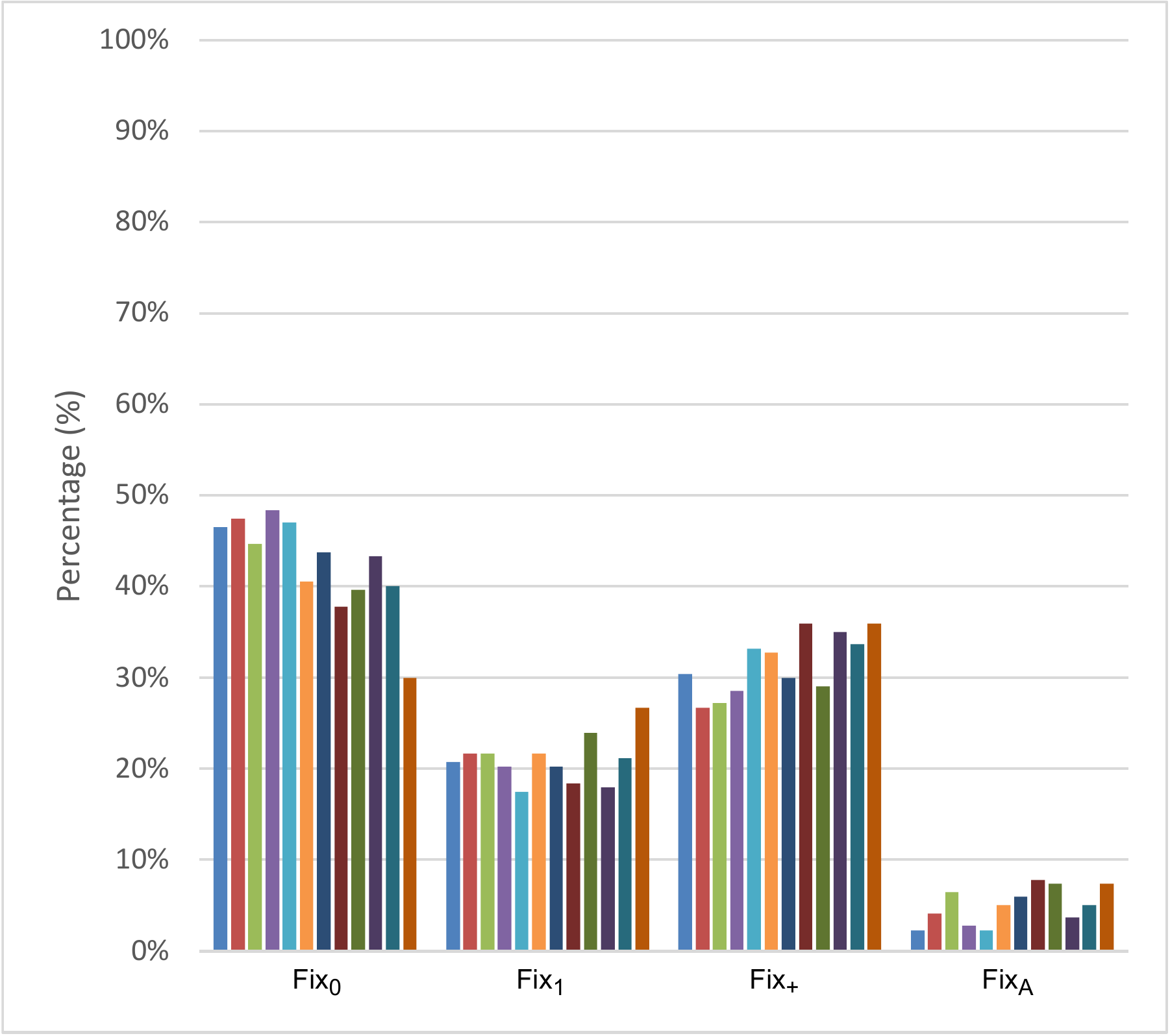}
        \caption{\footnotesize \% of Bugs with SYE}
        \label{fig:sye_apply_consistency}
    \end{subfigure}
    \hfill
    \begin{subfigure}[b]{0.33\textwidth}
        \includegraphics[width=\textwidth]{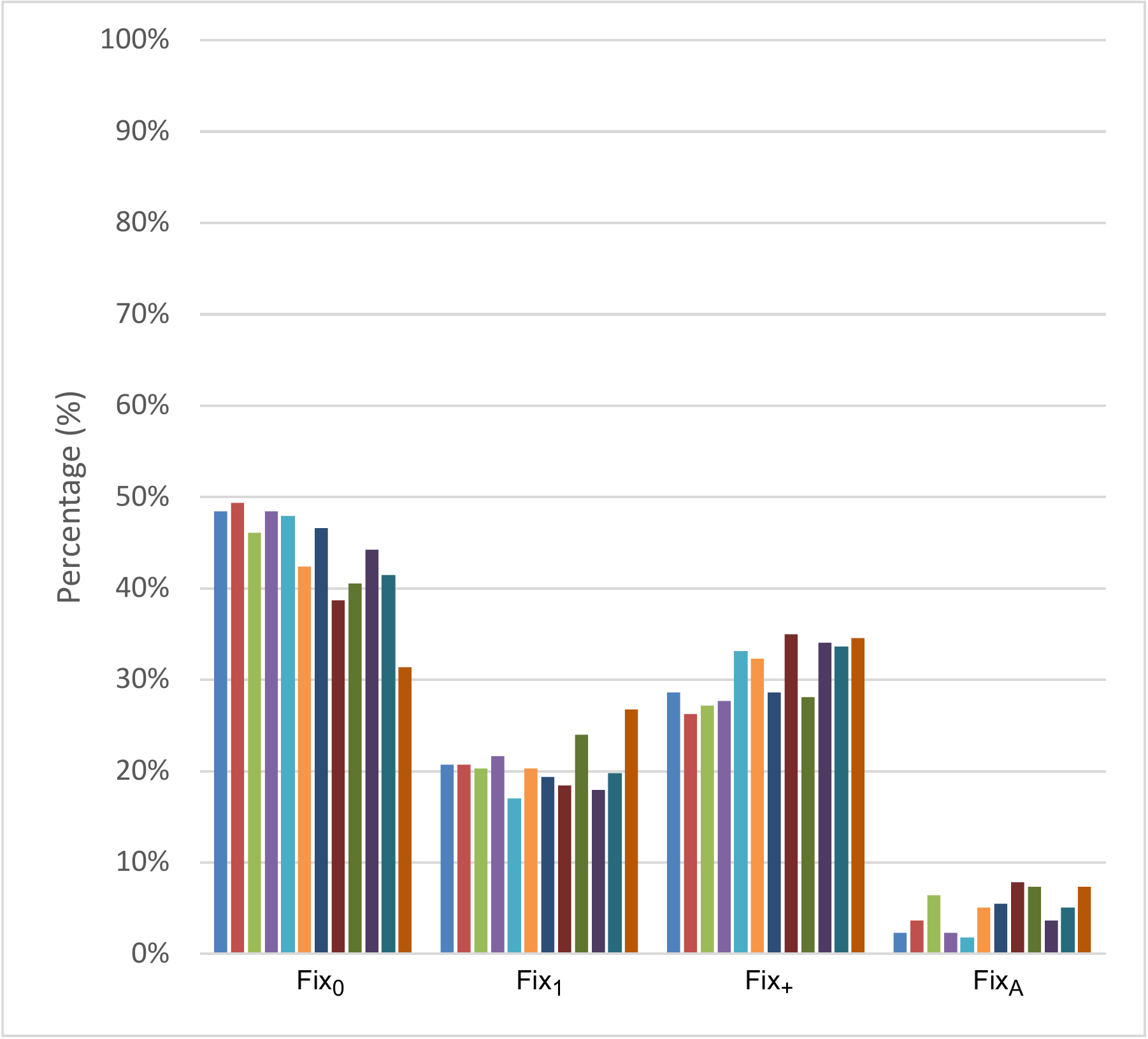}
        \caption{\footnotesize \% of Bugs with EM}
        \label{fig:em_apply_consistency}
    \end{subfigure}
  \caption{\add{\% of bugs with PP/SYE/EM under consistency-oriented multiple synthetic variants at the \texttt{Apply} layer.}}
  \label{fig:apply_consistency}
\end{figure*}

\add{
\subsubsection{\texttt{Apply}} \hfill\\
At the \texttt{Apply} layer, the focus shifts from source-level understanding to transferring bug-fixing knowledge (bug patterns in particular) into a new but still controlled context. 
In our current prototype, the main \texttt{Apply}-layer case study is realized through synthetic bug-pattern generation in newly constructed lightweight local contexts. The same-root-cause synthetic setting is used later as a controlled extension for interpreting cross-layer difficulty, rather than as the primary operationalization of the \texttt{Apply} layer itself. More generally, this layer can be expanded through the following dataset-construction strategies:

\begin{itemize}[leftmargin=*]
    \item \textbf{Cross-language or newer-language transfer}: Generates synthetic variants that preserve a similar bug pattern but re-express it in another programming language or in a newer version of the same language, thereby testing whether repair knowledge transfers beyond the original implementation setting.
    \item \textbf{Controlled synthetic re-instantiation consistency}: Creates multiple synthetic buggy functions that preserve the same bug pattern while varying the surrounding local implementation context, assessing whether the APR system maintains consistent repair behavior.

To systematically evaluate the consistency of the SUEs, we have established the following classification of the SUE behavior while fixing a particular bug instance from the original static benchmark under different synthetic instances. 

\begin{itemize}
    \item \emph{$FIX_0$} indicates that none of the synthetically generated bugs were fixed, suggesting the SUE cannot handle this bug type.

    \item \emph{$FIX_1$} indicates that only one of the synthetically generated bugs was fixed, reflecting limited capability.

    \item \emph{$FIX_+$} indicates that more than one, but not all, synthetically generated bugs were fixed, suggesting constrained capability.

    \item \emph{$FIX_A$} indicates that all synthetically generated variants produced the target repair outcome, indicating high consistency across the evaluated variants. 
    
\end{itemize}


\end{itemize}


\jiho{Suggest giving the denominators and construction protocol: \#of src bugs, variants per bug, prjs/contexts, eligibility/exclusions, generators/prompts, validation procedure, etc. This would help understand FIX percentages in fig 9 and 10 and support the response to E0.2/R2.2.}
\emph{Pilot Study}: For each of the 217 source bug IDs, we constructed five synthetic variants, yielding 1,085 variants in total. We followed the generation procedure described in Section~\ref{sec:framework} and used automated scripts to verify that the generated code and corresponding tests differed across the five variants associated with each source ID.

Figure~\ref{fig:apply_consistency} illustrates the consistency evaluation on the synthetic bug-pattern dataset for the \defectj-based synthetic bug-pattern consistency extension study. The PP results remain highly stable across repeated re-instantiations: $FIX_0$ ranges only from 0.00\% to 4.61\%, while $FIX_A$ ranges from 39.17\% to 80.65\%, and the combined $FIX_+ + FIX_A$ rate reaches 86.17\% to 100.00\%. The corresponding SYE and EM results are lower but still comparatively stable, with $FIX_+ + FIX_A$ ranging from 30.88\% to 43.77\% for SYE and from 29.95\% to 42.85\% for EM. These patterns suggest that, under lightweight synthetic local contexts, current APR systems can often maintain relatively consistent repair behavior across multiple re-instantiations of the same bug pattern, especially at the PP level.\respto{E0-2.4}\respto{R2-2.3}} 

\subsubsection{\texttt{Analyze}} \hfill\\
At the \texttt{Analyze} layer, the framework probes a deeper form of transfer. Here, the evaluation no longer centers on preserving the same surface buggy pattern, local implementation shape, or patch form as the original benchmark instance. Instead, the key requirement is to preserve the same underlying root cause or failure mechanism while embedding the bug into a substantially different context. In our current case studies, this is instantiated through one-to-one matched real-world cases. More broadly, this layer can be expanded through the following strategies:
\jiho{Analyze is defined by preserving the root cause/failure mechanism, but this bullet also permits “bug-fixing patterns,” which seem to overlap the Apply definition?}
\begin{itemize}[leftmargin=*]
    \item \textbf{Intra-project root-cause consistency}: within the same project, inject different code-level bug patterns that all preserve the same root cause, and evaluate whether the APR system remains stable across those alternative surface realizations.
    \item \textbf{Cross-project transfer consistency}: re-instantiate the same underlying failure root cause across multiple project contexts while allowing the surface buggy pattern, local implementation, and patch form to differ, and evaluate whether the APR system can maintain stable repair behavior under these contextual shifts.
\end{itemize}

\begin{figure*}[t]
  \centering
  \includegraphics[width=0.7\textwidth]{figure/legend1.pdf}

    \begin{subfigure}[b]{0.33\textwidth}
        \includegraphics[width=\textwidth]{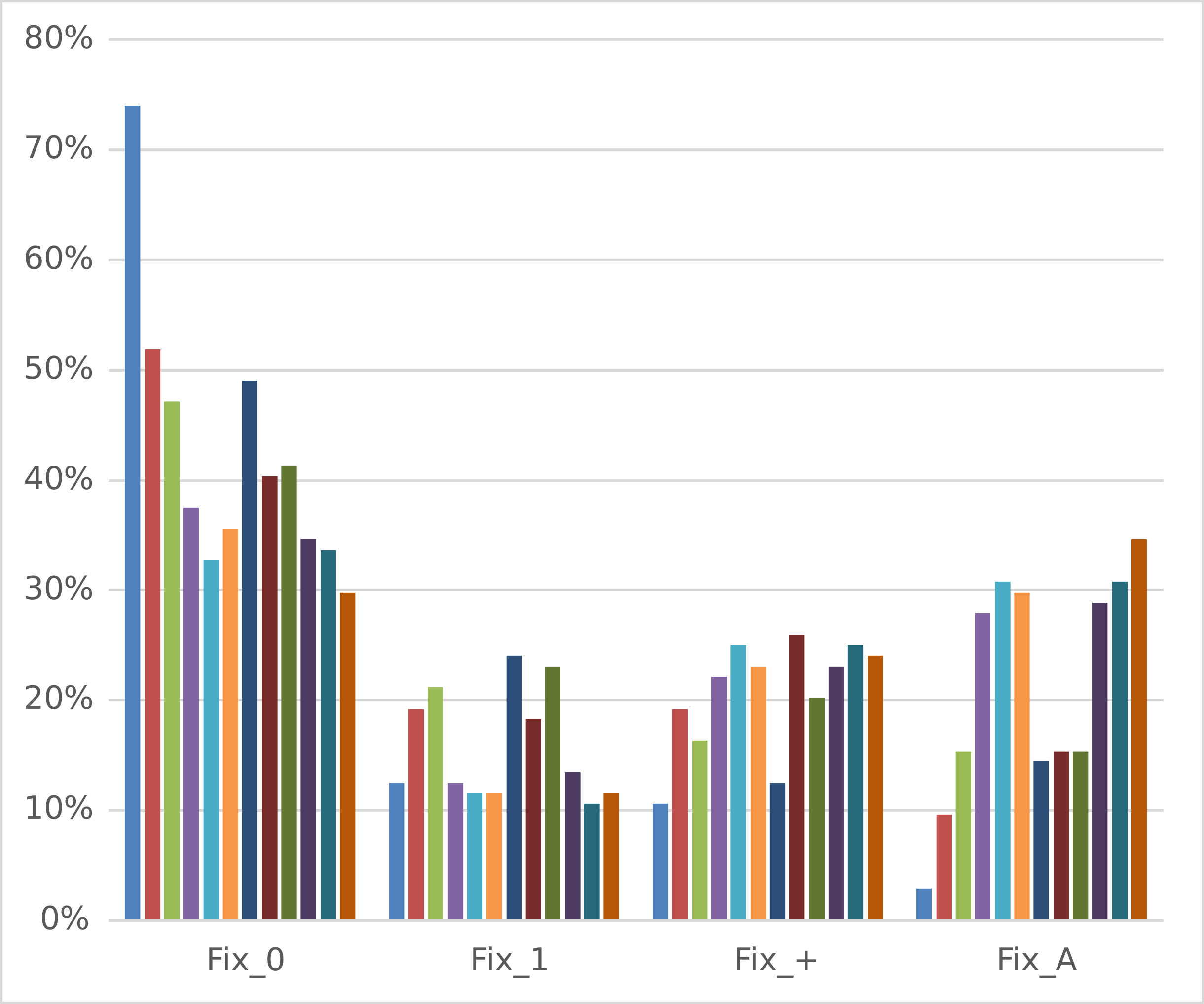}
        \caption{\footnotesize \% of Bugs with PP}
        \label{fig:pp_apply}
    \end{subfigure}
    \hfill
    \begin{subfigure}[b]{0.33\textwidth}
        \includegraphics[width=\textwidth]{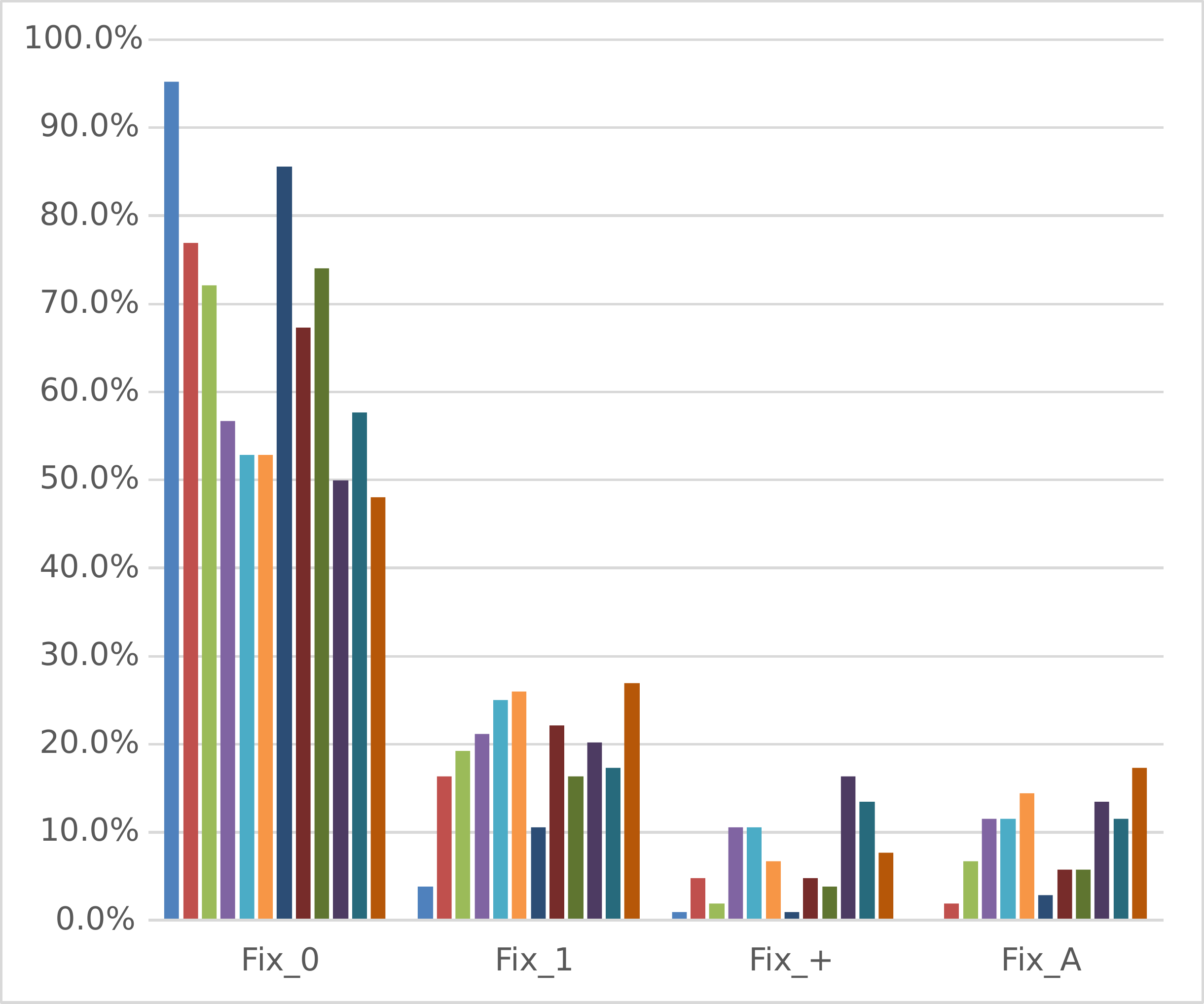}
        \caption{\footnotesize \% of Bugs with SYE}
        \label{fig:sye_apply}
    \end{subfigure}
    \hfill
    \begin{subfigure}[b]{0.33\textwidth}
        \includegraphics[width=\textwidth]{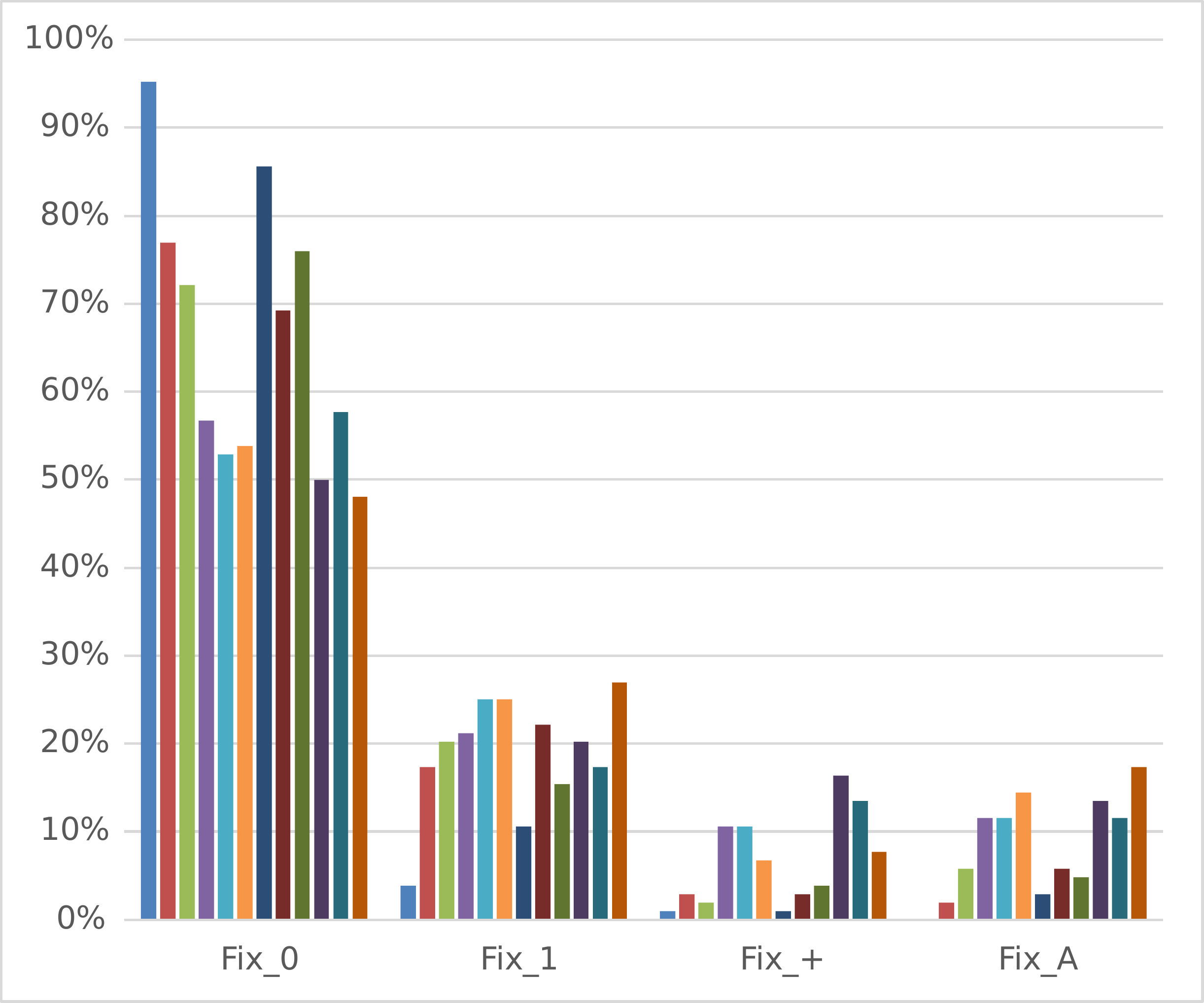}
        \caption{\footnotesize \% of Bugs with EM}
        \label{fig:em_apply}
    \end{subfigure}

  \caption{\% of bugs with PP/SYE/EM under cross-project consistency-oriented evaluation across 104 Defects4J source bug IDs, covering 261 validated real-world variants in total at the \texttt{Analyze} layer.}
  \label{fig:rq4}
\end{figure*}

\add{\emph{Pilot Study}: This analysis includes 261 validated real-world bug instances derived from 104 \defectj source bug IDs. Depending on the availability of suitable root-cause-matched code contexts, each source bug ID is associated with between one and five validated real-world variants. For source bug IDs with only one available real-world variant, a successful repair is classified as $FIX_A$ because all available variants were repaired. $FIX_1$ is reserved for source IDs with at least two available variants for which exactly one variant produces the target repair outcome. This variation is a consequence of the real-world search and validation process: some source bugs yield multiple valid project contexts, whereas others yield only one. For each source bug ID and SUE setup, the $FIX_0$, $FIX_1$, $FIX_+$, and $FIX_A$ categories summarize repair consistency across the real-world variants available for that source ID. Because the number of available variants differs across source IDs, the consistency categories are defined relative to each source ID’s available validated variants rather than to a fixed number of variants per bug.

Figure~\ref{fig:rq4} shows the consistency-oriented \texttt{Analyze}-layer evaluation under the cross-project root-cause setting, where the same underlying root cause is re-instantiated across broader real-world project contexts. Compared with the lighter-weight synthetic settings, consistency becomes substantially harder to maintain in this more realistic context. At the PP level, $FIX_0$ ranges from 29.81\% to 74.04\%, $FIX_A$ ranges from only 2.88\% to 34.62\%, and the combined $FIX_+ + FIX_A$ rate ranges from 13.46\% to 58.66\%. The SYE and EM results are even weaker: for both metrics, $FIX_0$ ranges from 48.08\% to 95.19\%, while the combined $FIX_+ + FIX_A$ rate reaches only 0.96\% to 29.81\%. These results indicate that once the same root cause must be preserved under broader real-world contextual variation, current APR systems often fail to maintain stable repair behavior across variants. In other words, relative to the synthetic \texttt{Apply} and \texttt{Root Cause} settings, these results are consistent with the interpretation that realistic contextual complexity introduces additional difficulty beyond root-cause preservation; however, the comparison remains descriptive rather than causal. \respto{R2-2.4}}

\subsubsection{\texttt{Evaluate} and \texttt{Create}} \hfill\\
For the remaining Bloom layers, although our prototype does not operationalize them, \tool can accommodate future evaluations as follows:

\noindent \textbf{The \texttt{Evaluate} Layer}: Current APR benchmarks are dominated by pass/fail automated assessment, which limits the ability to evaluate whether a system can critically compare multiple plausible repairs and justify why one solution is preferable to another. This layer can be operationalized along the following directions:

\begin{itemize}[leftmargin=*]
    \item \textbf{Stronger PP evaluation}: assess plausible patches using better testing, stronger coverage, or richer criteria beyond functionality, such as code quality, design quality, or architecture-level constraints.
    
    \item \textbf{Repair validation without pre-existing tests}: construct tasks in which test cases are not provided in advance, so that the system must generate tests to verify whether its own fixes are acceptable.
\end{itemize}


\noindent \textbf{The \texttt{Create} layer}: the benchmark can be extended toward genuinely new repair synthesis. This layer can be operationalized through the following directions:

\begin{itemize}[leftmargin=*]
    \item \textbf{Unseen bug-fixing patterns or workflows}: evaluate whether APR systems can repair bugs that require reasoning strategies not directly recoverable from familiar benchmark families.
    \item \textbf{Newly curated or compound bugs within the same project context}: keep the project environment fixed, but introduce newer or compound bugs so that the system must reason under a stable context while addressing unfamiliar bug scenarios.
\end{itemize}

In summary, the preceding discussion and pilot studies demonstrate that the primary contribution of \tool lies not in any single mutation recipe, but in the structured, layered methodology itself. The \defectj and \SWE case studies serve as a proof of concept that this framework can be concretely instantiated to expose critical limitations in LLM-powered APR systems. Furthermore, the broader design space summarized in Table~\ref{tab:layer_generalization} illustrates that each layer accommodates multiple valid realizations, a flexibility essential for future benchmark evolution. Rather than treating a specific transformation or benchmark adaptation as definitive, we view this framework as a principled foundation for organizing progressively richer, more realistic evaluations of code understanding, controlled transfer, root-cause reasoning, repair assessment, and creative synthesis.

\subsection{Using Bloom's Taxonomy to Evaluate LLM-powered SE Systems}

In this paper, we emphasize the urgent need to update static benchmarks, which make up the majority of current SE benchmarks, into dynamically evolving ones to more faithfully assess and improve the capabilities of LLM-powered SE solutions. Instead of focusing solely on eliminating overlap with LLM training data, adapting static benchmarks using Bloom's Taxonomy offers a promising path for comprehensive evaluation. While our framework is demonstrated on APR benchmarks (\defectj and \SWE), it is also applicable to many other SE use cases, such as test case generation (e.g., TestGenEval~\cite{jain2024testgeneval}) and vulnerability detection (e.g., Vul4J~\cite{bui2022vul4j}, Big-Vul~\cite{fan2020ac}).

\section{Threats to Validity}
\label{sec:threats}

\noindent \textbf{Internal Validity}:
While evaluating the capabilities of LLM-powered APR solutions, we identified two primary sources of potential confounding factors: 
(1) \emph{Possible Data Contamination}: we established a memory-permissive reference baseline by evaluating SUEs on the original \defectj benchmark, which likely overlapped with the training datasets from the underlying LLMs due to its long-standing public availability. We then evaluated SUEs on dynamically generated bugs that are textually and contextually distinct but semantically similar to the original bugs. By comparing evaluation results across these bug datasets, we examined performance changes that may be associated with benchmark familiarity and possible data contamination. 

(2) \emph{Bug Characteristics}: To ensure fair comparison across different \tool layers,  the dynamically generated bugs were crafted to preserve the layer-specific bug characteristics defined in Section~\ref{sec:framework}, such as the original repair target at the \texttt{Understand} layer, a similar buggy behavior pattern at the \texttt{Apply} layer, and the same underlying root cause at the \texttt{Analyze} layer. These layer-specific invariants support cross-layer analysis and interpretation of the evaluation results. 

\noindent \textbf{External Validity}:  
\emph{Choice of LLMs and Agents}: 
This paper presents a framework for transforming static SE benchmarks into dynamic benchmarks structured by Bloom's Taxonomy. \add{To demonstrate the effectiveness and generalizability of our framework, we conducted two case studies: (1) evaluating ChatRepair and CigaR on the Java-based \defectj benchmark, and (2) evaluating SWE-agent on the Python-based \SWE benchmark. Many LLM-powered APR systems, such as OpenHands~\cite{wang2024openhands} and AutoCodeRover~\cite{zhang2024autocoderover}, have been introduced recently. However, these agents' implementations are often tightly coupled with a specific benchmark~\cite{kapoor2024ai}, hindering direct cross-comparison across different benchmarks or evaluation criteria.
Even with the current number of SUEs and LLMs, conducting the evaluation required more than six months and incurred over \$42,800 in billed \tool project costs, despite extensive pipeline optimization and parallelization.\respto{R2-1.1} \respto{R1-6.3}}
We emphasize that the goal of this paper is not to benchmark exhaustively but to demonstrate the feasibility and preliminary cross-benchmark applicability of \tool. 

\noindent \emph{Generalizability of Empirical Findings}: Since our evaluation focuses specifically on single-function bugs in Java and Python, the findings may not generalize to other programming languages (e.g., JavaScript or Haskell) or bug types (e.g., performance or security bugs). Additionally, results from this case study may not extend to non-LLM-powered APR systems (e.g., rule-based systems) or LLM-powered systems operating in other SE tasks (e.g., code review or requirements analysis).

\noindent \textbf{Construct Validity}:  
\emph{Evaluation Metrics}:
Our study adopts PP, SYE, and EM as the primary evaluation metrics, consistent with prior work in this area~\cite{10.1145/3293882.3330577, capgen, yuan2018arja, jiang2018shaping, cure, sequencer, selfapr, rewardrepair, bouzenia2024repairagent, yangmorepair, ahmed2023better, li2025hybrid}. While these metrics help estimate patch quality, they have limitations: passing test cases does not guarantee correctness, and multiple valid fixes may exist for the same bug. Further research is needed to develop automated techniques capable of reliably assessing the correctness of patches generated by LLM-powered APR systems.

\noindent \emph{Identifier names}: Variable and function names often carry semantic meaning. Prior studies~\cite{WuISSTA2023,zhang2025UnseenHorizons,wang2022recoderobustnessevaluationcode} show that meaningful variable and/or function names aid LLMs in understanding code better. Our program perturbations (\texttt{Hash-based renaming} and \texttt{Rephrasing-based renaming}) at the \texttt{Understand} layer intend to deliberately test whether APR systems rely on superficial identifiers versus deep reasoning about the code logic. Function name rephrasing is content-based, while variable name rephrasing is self-referential. These transformations were designed to preserve runtime behavior, the original bug logic, and the repair target, while deliberately altering or removing identifier-level semantic cues. We manually verified that these changes did not alter the intended meaning when developing our framework. The goal here is to evaluate whether APR systems can comprehend code logic and apply correct bug-fixing patterns for familiar bugs.

\noindent \emph{Manual Evaluation}: The manual-validity assessment was conducted by two authors who divided the twelve setups equally, with each sampled case assessed by one assigned author using the same evaluation criteria. Because the cases were not independently double-coded, we do not report inter-rater agreement, and reviewer subjectivity may affect the sample-based manual-validity estimates.


\section{Conclusion}
\label{sec:conclusion}


\add{
LLM-powered APR systems have shown promising results in recent years, yet their evaluation still relies heavily on static benchmarks such as \defectj and \SWE. These benchmarks suffer from critical limitations, including the risk of data contamination and a lack of contextual diversity needed to robustly assess reasoning and transfer capabilities. In this paper, we propose \tool, a dynamic evaluation framework grounded in Bloom's Taxonomy. While Bloom's Taxonomy contains six layers, the current version of \tool operationalizes the first four layers (\texttt{Remember}, \texttt{Understand}, \texttt{Apply}, and \texttt{Analyze}) to provide a structured evaluation ladder for LLM-powered APR systems.

By applying \tool to ChatRepair and CigaR on adapted \defectj, and to SWE-agent on adapted \SWE across multiple LLM backbones, we find that while these systems perform strongly at the memory-permissive \texttt{Remember} layer, we observe a non-monotonic performance pattern across the four operationalized layers. Specifically, lexical perturbations at the \texttt{Understand} layer substantially degrade repair performance. The synthetic bug-pattern transfer at the \texttt{Apply} layer yields high PP, but SYE and EM remain substantially lower, indicating that test-passing repairs are easier to obtain than faithful repairs in these controlled settings.
Finally, on \defectj, the \texttt{Analyze} layer is the lowest-performing setting across the evaluated metrics. On \SWE, \texttt{Analyze} remains substantially more difficult than the \texttt{Apply} setting, but it is not uniformly the lowest-performing layer across all models and metrics. 
These findings reveal a clear gap between benchmark-level success and robust, context-sensitive repair in substantially altered or realistic project environments.


Our study highlights both the limitations of current LLM-powered APR systems and the limitations of the static benchmarks used to assess them. More broadly, it shows that existing static APR benchmarks can be systematically transformed into \tool-style dynamic evaluations, enabling progressively richer assessment of code understanding, controlled transfer, and root-cause-level reasoning. We hope this framework encourages more robust and trustworthy evaluation of AI-driven SE systems.
}


\bibliographystyle{ACM-Reference-Format}
\bibliography{reference}

\end{document}